\documentclass[journal]{IEEEtran}

\ifCLASSOPTIONcompsoc
\else
\fi

\usepackage{graphicx}
\usepackage{cite}
\usepackage[cmex10]{amsmath}
\usepackage{amssymb}
\usepackage{array}
\usepackage{subfigure}
\usepackage{url}
\usepackage{multirow}
\usepackage{algorithmic}
\usepackage{algorithm}

\usepackage{cuted, flushend}
\usepackage{color}

\begin{document}
%
% paper title
% can use linebreaks \\ within to get better formatting as desired
\title{Load Balancing in Two-Tier Cellular Networks with Open and Hybrid Access Femtocells}

\author{Dongmyoung Kim,~%\IEEEmembership{Non-member,}
        Taejun Park,~%\IEEEmembership{Non-member,}
        Hyoil Kim,~%\IEEEmembership{Member,~IEEE}
        and~Sunghyun~Choi%,~\IEEEmembership{Fellow,~IEEE}% <-this % stops a space
}

% The paper headers
\markboth{\tiny THIS WORK HAS BEEN SUBMITTED TO THE IEEE FOR POSSIBLE PUBLICATION. COPYRIGHT MAY BE TRANSFERRED WITHOUT NOTICE, AFTER WHICH THIS VERSION MAY NO LONGER BE ACCESSIBLE }%,~Vol.~X, No.~X, OCTOBER~2012}%
{Shell \MakeLowercase{\textit{et al.}}: Bare Demo of IEEEtran.cls for Computer Society Journals}

% for Computer Society papers, we must declare the abstract and index terms
% PRIOR to the title within the \IEEEcompsoctitleabstractindextext IEEEtran
% command as these need to go into the title area created by \maketitle.

% make the title area
\maketitle

%%% 01 Abst begin %%%

\begin{abstract}
Femtocell base station~(BS) is a low-power, low-price BS based on cellular communication technology. It is expected to become a cost-effective solution for improving the communication performance of indoor users, whose traffic demands are large in general.
There are mainly three access strategies for femtocell, i.e., closed access, open access and hybrid access strategies.
While it has been generally known that open/hybrid access femtocells contribute
more to enhancing the system-wide performance than closed access femtocells,
the operating parameters of both macro and femtocells should be carefully
chosen according to the mobile operator's policy, consumer's requirements, and so on.
We propose long-term parameter optimization schemes,
which maximize the average throughput of macrocell users while guaranteeing some degree of benefits to femtocell owners. To achieve this goal, we jointly optimize the ratio of dedicated resources for femtocells as well as the femtocell service area
in open access femtocell networks through the numerical analysis.
It is proved that the optimal parameter selection of open access femtocell is
a convex optimization problem in typical environments.
Then, we extend our algorithm to hybrid access femtocells where
some intra-femtocell resources are dedicated only for femtocell owners while remaining resources are shared with foreign macrocell users.
Our evaluation results show that the proposed parameter optimization
schemes significantly enhance the performance of macrocell users thanks to the large offloading gain. The benefits provided to femtocell users are also adaptively maintained according to the femtocell users' requirements.
The results in this paper
provide insights about the situations where femtocell
deployment on dedicated channels is preferred to the co-channel deployment.

\end{abstract}

%%% 01 Abst end %%%


%\IEEEdisplaynotcompsoctitleabstractindextext
% \IEEEdisplaynotcompsoctitleabstractindextext has no effect when using
% compsoc under a non-conference mode.

%\IEEEpeerreviewmaketitle

%%% 02 Intro begin %%%
\section{Introduction}
\label{sec:intro}

\IEEEPARstart{A}{femtocell} base station
%KHI:
abbreviated as femto BS or fBS
is a small BS  with low transmission power and low cost.
Femtocells can be installed by the end users to enhance the cellular networking performance at home,
of which traffic is transported via an Internet backhaul such as Digital Subscriber Line~(DSL) or cable modem.
Two-tier cellular networks, consisting of a conventional macrocell network
and underlaying short-range femtocells,
have received considerable attention from industry and academia as an efficient solution
to deal with the exploding demand for wireless data communication.
The femtocell technology has an advantage over other competing indoor wireless communication
technologies, thanks to its high capacity and backward compatibility with existing cellular technologies.
The history, current status of market and technology,
research issues, and future expectation of femtocell technology
are well summarized in~\cite{jsac12andrews}.

There are mainly three strategies for a femtocell access, namely,
closed access, open access, and hybrid access strategies.
A femtocell in a closed access mode can only be accessed by
authorized femtocell users.
On the other hand, if the owner of a femtocell installs it with the
open access mode, any macrocell user might access the
femtocell. Some previous researches~\cite{iccs09lopez,globecom10xia,icc11jo}
have shown that the deployment of open access femtocells can improve the system-wide performance by transferring some of the traffic loads in congested macrocells to the femtocells.
%KHI:
Hybrid access mode is a compromise between closed and open access, where
an fBS allows arbitrary nearby users to
access it like open access mode but the subscribed femtocell owners can be prioritized over
unsubscribed users. The prioritization can be implemented by using  various
vendor-specific mechanisms.
%
%PTJ:
%In 3GPP LTE standards,
In 3GPP Release~8 specification~\cite{3gppTS36.3.8},
only closed and open access modes are supported for femtocells
%PTJ:
%in 3GPP Release~8 specification~\cite{3gppTS36.3.8}
while the hybrid access mode has been added in 3GPP Release~9 specification~\cite{3gppTS36.3.9}.
Therefore, considering both the open and hybrid access strategies is important.
In this paper, we numerically analyze and optimize the performance of both open and hybrid access femtocell
networks. We propose the load balancing schemes which properly balance the traffic loads in macrocells and femtocells. Our load balancing schemes aim at maximizing the system-wide performance,
i.e., the average throughput of the users communicating with the macrocells,
in two-tier cellular networks with open or hybrid access femtocells,
%KHI:
while guaranteeing some benefits of the femtocell owners
such that femtocell users can always achieve larger throughput than macrocell users.
%KHI:
Such an approach not only improves the macrocell user's performance via traffic offloading from macrocell to femtocell, but also promotes the deployment of femtocells via the guaranteed benefit to the femtocell owners.

%KHI:
In our proposed framework, we strike a balance between a macrocell and femtocells by controlling the service area of femtocells because the amount of traffic loads is dominated by the number of associated users.
In order to maximize the offloading efficiency, orthogonal deployment --- in which
%KHI:
the whole wireless resources are divided into two parts, one dedicated to femtocells and the other reserved for a macrocell --- is considered, and we jointly optimize the amount of resources dedicated to the femtocells and the service area of femtocells.
The optimization problem is first studied for the open access case, and then extended to the hybrid access femtocell networks, where a variable portion of intra-femtocell resources is exclusively used by femtocell owners while remaining resources are shared with macrocell users associated with the femtocell.

The contributions of this paper are summarized as follows:
\begin{enumerate}
\item We numerically analyze the performance of macrocell and femtocell users in two-tier cellular networks
where open and hybrid access fBSs are deployed in an un-planned manner.
\item Multiple essential parameters in the open and hybrid access femtocell networks are jointly optimized
to enhance the performance of both the macrocell and femtocell users.
\item Our results show that the orthogonal spectrum dedication for open and hybrid access femtocell can be more beneficial than co-channel deployment in some aspects. The conditions that femtocell deployment on dedicated channel can be preferred are also discussed.
\item It is proved that the joint optimization of the amount of dedicated resources and the service area of
femtocells is a convex optimization problem in typical environments.
\end{enumerate}

The rest of the paper is organized as follows. Section~\ref{sec:related} presents the related work. In Section~\ref{sec:system_model}, we introduce the system model and our load balancing
problem formulation in the open and hybrid access femtocell based two-tier cellular networks.
We analyze the average throughput of each type of users in Section~\ref{sec:analysis} to complete the problem formulation.
In Section~\ref{sec:optimization_open}, we obtain the optimal
parameters of open access femtocell networks and
some theorems for optimal parameter selection are discussed. By extending the
optimization framework of open access femtocell networks,
system parameters of hybrid access femtocell are also optimized
in Section~\ref{sec:optimization_hybrid}.
In Section~\ref{sec:evaluation}, our proposed schemes are evaluated based on both numerical analysis and computer simulations.
Finally, we conclude the paper in Section~\ref{sec:conclusion}.

%%% 02 Intro end

%%% 03 Related Work begin %%%
\section{Related Work}
\label{sec:related}

Many resource allocation schemes have been proposed for the two-tier cellular networks, but most of the proposed schemes are heuristic or locally optimized schemes~\cite{coml08guvenc,globecom08choi,icc09claussen,mobihoc09sundaresan,
globecom09bai,twc10jo,icc11kaimaletu,icc11park,mobicom11arslan,icc11hatoum}.
On the other hand, some previous work conducts optimization based on
full channel information~\cite{wiopt09han,pimrc09kim,twc11cheng} or %,jsac11tan} or
game theoretic model~\cite{pimrc10bennis,icc11lin,icc11bennis,tvt11ko}.
Therefore, the previous researches are different from our work which optimizes
the system-wide performance based on the long-term system information
of the two-tier cellular networks where fBSs are deployed in an unplanned manner.
Our long-term parameter optimization framework is not incompatible with but complementary to
the short-term resource allocation schemes
in the sense that the long-term optimization framework can provide good guidelines
for the parameter configuration considering the system-wide average performance.

Recent papers~\cite{twc10jo,tcom09chandrasekhar} are the most relevant previous work in the literature.
Coverage control schemes have been proposed in~\cite{twc09Chandrasekhar_coverage,twc10jo}.
The authors in~\cite{twc10jo} proposes an adaptive transmit power control scheme to control
the shape and the size of femtocell coverage.
In closed access femtocell networks, which is the system model of the above mentioned papers,
the objective of coverage control is minimizing the interference leakage from fBSs to the outdoor macrocell region
while the expected service area for the subscribed users is guaranteed.
Service area adaptation in this paper is different
because the strong signal received from a femtocell is
considered a good serving signal in open and
hybrid access femtocell networks, which allow the macrocell users to access them.

In~\cite{tcom09chandrasekhar},
the authors  propose a bandwidth division scheme in the two-tier cellular networks composed of the closed access femtocells. The objective and constraints of~\cite{tcom09chandrasekhar} are different from our work %KHI:
since ours
utilizes the traffic offloading gain in the open and hybrid access femtocell networks.
Furthermore, we optimize some other control parameters, such as target service area of a femtocell and intra-femtocell resource dedication ratio for a femtocell owner, together with the bandwidth division ratio to enhance the system performance.
%%% 03 Related Work end %%%

%%% 04 System Model begin %%%
%PTJ:
\section{System Model and Our Framework}
\label{sec:system_model}
\subsection{System Model}
\label{subsec:femto_model}

We assume a single circular macrocell region with
%KHI:
the radius of $D_m$ and the area of $A_m = \pi {D_m}^2$,
where a macrocell BS (mBS) is located at the center of the circular region.
%KHI:
Multiple
fBSs are randomly distributed within the macrocell region according to
a homogeneous Spatial Poisson Point Process~(SPPP)~\cite{book93kingman} with intensity $\lambda_f$.
In an SPPP, the number of points in a given region follows Poisson random variable
with the mean of $\lambda A$, where $\lambda$ and $A$ are the intensity of the points
and the area of the given region, respectively.
We assume that each fBS is owned by a femtocell mobile station~(fMS),
and the fBS is located at the center of the indoor circular home region
with the radius of $D_{h}$.
%KHI:
The fMS
of each fBS is randomly located within the
circular home region, and the indoor home area and outdoor area are partitioned with a wall.

%PTJ:
%KHI:
It is possible that the fBSs following SPPP are located closer than $2D_h$ to each other thus resulting in overlapping home areas.
Although this should not happen in practice, we argue that the proposed SPPP model still well captures the reality because it is highly unlikely to have such overlapping fBSs with the practical range of $\lambda_f$ and $D_h$.
Suppose we denote by $P_{overlap}$ the probability that two or more fBSs overlap with each other. Since $P_{overlap}$ is identical to the probability that two or more fBSs exist in the area of $\pi (2 D_h)^2$, we have
\begin{eqnarray*}
P_{overlap} &=& 1 - e^{-\lambda_f \cdot 4 \pi {D_h}^2} - \lambda_f \cdot 4 \pi {D_h}^2 \cdot e^{-\lambda_f \cdot 4 \pi {D_h}^2}\\
&=& \begin{cases}
    0.079, & \textrm{in US},\\
    %{\color{red}??}, & \textrm{in UK},\\
    0.081, & \textrm{in Korea},
\end{cases}
\end{eqnarray*}
\noindent where the values of $\lambda_f$ and $D_h$ are obtained from {\cite{census2010_kr,census2010_us}}.
Hence, we henceforth assume that there exists only one fBS per home, i.e., the distance between any two fBSs is larger than $2D_h$.
In Section~\ref{sec:evaluation}, it will be verified that such an approach approximates typical environments with a reasonably small analytical error (less than $1$~\%) as shown in Fig.~\ref{fig:validation}.
Note that in our simulation, we generate fBSs according to SPPP and drop a new fBS located closer than $2D_h$ to any of the previously-generated fBSs.

Similarly to fBSs, macrocell users are randomly distributed according to an SPPP
with intensity $\lambda_u$ in the whole macrocell area.
Because some macrocell users can associate with a nearby femtocell
in the open and hybrid access femtocell networks,
the macrocell users are categorized into
%KHI:
two types,
i.e., macrocell mobile station~(mMS) and open access mobile station~(oMS).
We refer to the macrocell user who associates with mBS as mMS, while
the macrocell user who associates with an fBS thanks to the open or hybrid  access
policy is referred to as oMS.

The downlink channel gain between a BS and an MS is characterized by a pathloss and fading.
When the distance between a transmitter and a receiver is $d$, the channel
gain of the link is modeled by $\Psi\left( {Zd} \right)^{ - \alpha }$, where $\alpha$ is the pathloss exponent and $Z$ represents a fixed loss
which is dependent on the type of the link.
Different $Z$ and $\alpha$ values, i.e., $Z_1$ to $Z_5$ and $\alpha_1$ to $\alpha_5$, are
defined for different types of links as shown in Table~\ref{table:pathloss}.
$\Psi  \sim \exp \left( 1 \right)$ is a
Rayleigh fast fading component which has a unit average power.

We consider multiple discrete transmission rates, where the rate is adaptively determined according
to the Signal to Interference plus Noise Ratio~(SINR) value at the receiver.
%KHI:
Rate index $l \in [1,L]$ corresponds to the case when the SINR lies in $[\Gamma_l, \Gamma_{l+1})$, where $\Gamma_{L + 1} = \infty$.
and the spectral efficiency of
%KHI:
rate index $l$
is modeled as the following based on the variable rate
M-QAM transmissions:
\begin{equation}
\label{eq:rate_eq}
b_l  = \log _2 \left( {1 + \frac{{\Gamma _l }}{G}} \right),
\end{equation}
where $G$ denotes Shannon Gap introduced in~\cite{tcom97goldsmith}.
The specific rate set used in the simulations are summarized in Table~\ref{table:rate_sets}.
\begin{table}
 \centering
 \caption{Transmission rate set}\label{table:rate_sets}
 \begin{tabular}{|c||c||c|}
 \hline
 Rate index,  $l$ & Spectral efficiency, $b_l$ & SINR region (dB)  \\
                  & (bps/Hz) &    \\
 \hline\hline
 1 & 0.4922  & $[-4, 0)$\\ \hline
 2 & 1.3889 & $[ 0, 4)$\\ \hline
 3 & 2.8962 & $[ 4, 8)$\\ \hline
 4 & 4.7364 & $[ 8, 12)$\\ \hline
 5 & 6.6885 & $[ 12, 16)$\\ \hline
 6 & 8.6711 & $[ 16, \infty)$\\ \hline
 \end{tabular}
\end{table}
Furthermore, Table~\ref{table:parameters} provides the definitions and default values
of all notations frequently used in this paper.

\begin{table*}[ht]  \centering
  \caption{Definition of parameters and default values}\label{table:parameters}
  \begin{tabular}{|c||c||c|}
  \hline
  Symbol & Description & Default value  \\
  \hline\hline
  $D_m$, $A_m$ & Radius and area of a macrocell region  & $800$~m, $\pi D_m^2$\\ \hline
  $D_h$, $A_h$ & Radius and area of a home region  & $20$~m, $\pi D_h^2$\\ \hline
%  $D_{\max}$ & Maximum service radius of a femtocell & N/A \\ \hline
  $f_c$ & Carrier frequency & $2000$~MHz\\ \hline
  $W$ & System bandwidth & $5$~MHz\\ \hline
  $P_N$ & Noise power density & $-174$ dBm/Hz\\ \hline
  $P_m$ & Macrocell transmit power density & $46/W$ dBm/Hz\\ \hline
  $P_f$ & Femtocell transmit power density & $23/W$ dBm/Hz\\ \hline
  $WL$ & Wall penetration loss & $10$ dB\\ \hline
  $\Psi$ & Rayleigh fading component & N/A \\ \hline
  $\alpha$, $Z^{\alpha}$& Pathloss exponent and fixed loss value & See Table~\ref{table:pathloss} \\ \hline
  $b_l, \Gamma_l$ & Spectral efficiency and SINR threshold for rate index $l$& See Table~\ref{table:rate_sets}\\ \hline
  $\overline N_f,\lambda_f$ & Average number and intensity of fBSs in a macrocell area & $30$, $30/A_m$ \\ \hline
  $\overline N_{u}, \lambda_{u}$ & Average number and intensity of macrocell users (mMSs + oMSs)   & $200$, $200/A_m$\\ \hline
  $\rho$ & Ratio of femtocell resources to the whole bandwidth, $\rho \in \left[0,1\right]$& N/A \\ \hline
  $d_f$, $D_{\max}$ & Service radius  of a femtocell region and maximum value of $d_f$, respectively  & N/A \\ \hline
  $x$ & Average service area  of a femtocell region, $x \in \left[X_{\min},X_{\max}\right]$ & N/A \\ \hline
  $\theta$ & Resource usage probability in OA/HA-Thin, $\theta \in \left[0,1\right]$& N/A \\ \hline
  $\beta$ & Ratio of resources reserved for an fMS to the whole resources of a hybrid access fBS, $\beta \in \left[0,1\right]$ & N/A \\ \hline
  $\overline T_f$, $\overline T_m$, $\overline T_{o}$ & Average throughput of fMS, mMS, and oMS  & N/A \\ \hline
  $M$ ($K$) & Required ratio between average throughput of fMS (oMS) and mMS   & $10$, $1$\\ \hline
  $O_{\max}$  & Maximum average outage rate allowed for an oMS   & $0.15$ \\ \hline
  \end{tabular}
\end{table*}

In this paper, we
consider a fully loaded network environment where the BSs always have packets to transmit.
Furthermore, we assume that the scheduler in the mBS or the open access fBS
allocates the resource blocks in a round-robin manner
so that the whole wireless resources in the cell are equally distributed to
the associated users
%KHI:
in a long-term.
The basic level of fairness, i.e., intra-cell resource fairness,  is guaranteed
from this assumption.
In the hybrid access fBS, the scheduler reserves some amount of resources
for the fMS while the remaining resources are allocated using a round-robin manner.
We assume that the transmission power spectral densities, i.e., power per Hz,
of the mBS and fBSs are fixed as $P_m$ and $P_f$, respectively, and the noise power density is given by $P_N$.

\subsection{Our Framework}
\label{sec:modeling}
\label{subsec:updated_formulation}
\begin{figure}
\begin{center}
\includegraphics[width=0.30\textwidth]{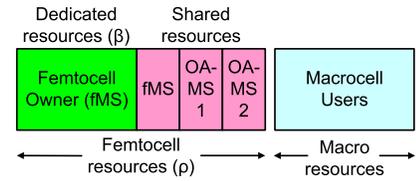}
\caption{Resource dedication in hybrid access femtocells ($\rho$ and $\beta$).}
\label{fig:hybrid_model}
\end{center}
\end{figure}

In this section, we introduce our load balancing framework
for open and hybrid access femtocell networks.
As shown in Fig.~\ref{fig:hybrid_model},
our schemes divide the whole available resources into two orthogonal sub-parts, and the two parts are dedicated to the macrocell and femtocells, respectively.
We refer to the ratio of resources dedicated to the femtocells
among the whole available resources as $\rho  \in \left[ {0,1} \right]$,
and we optimize $\rho$ to properly balance the traffic loads in the macro and femtocells.
We assume that the wireless resources can be divided either in time domain or frequency domain or both.

%KHI:
Our framework allocates the separate resources to the femtocells because
the resource separation not only limits the side effect to the existing macrocell users due to the femtocell deployment but also maximizes the offloading capability in the two-tier cellular networks by increasing the maximum cell coverage.
Though it has been generally said that open access femtocell networks prefer the co-channel deployment
option, the results of this paper show that
%KHI:
algorithms based on separate bandwidth can be more beneficial
in some aspects thanks to the enhanced
offloading gain and the increased flexibility to control the performance of fMSs, mMSs, and oMSs.
%KHI:
Appendix~\ref{sec:orthogonal} summarizes the benefits and preferred conditions of orthogonal deployment.

A hybrid access fBS allows the macrocell users to
access like open access fBS, but the intra-cell resource scheduler gives
priority to the fMS. It is different from the intra-cell resource allocation
policy of the open access femtocell where the resources are equally allocated to all the users
without distinguishing the fMS from the other macrocell users.
As shown in Fig.~\ref{fig:hybrid_model}, we assume that $\beta$ fraction of intra-femtocell resources are
dedicated to the fMS, and the remaining resources are equally shared by the
fMS and oMSs. Open access femtocell is a special case of hybrid access femtocell where $\beta=0$.

\begin{figure}
\begin{center}
\includegraphics[width=0.40\textwidth]{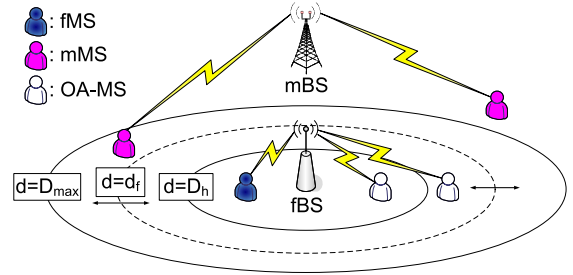}
\caption{Femtocell service radius ($d_f$) and user associations.}
\label{fig:ho_model}
\end{center}
\end{figure}
Our load balancing scheme in the open and hybrid access femtocell networks
jointly optimizes the average service area of a femtocell as well as the
amount of bandwidth dedicated to femtocells and the amount of intra-femtocell
resource dedicated to an fMS.
%
%KHI:
We
optimize the service coverage of a femtocell because
the cell selection based on the strongest RSS value is not efficient
to promote the load balancing.
Fig.~\ref{fig:ho_model} describes the system model for the service coverage optimization.
We refer to the service radius of a femtocell as $d_f$, and the femtocell and macrocell
users who are located closer than $d_f$ associate with the femtocell rather than
the macrocell.  Each fBS is required to fully cover the indoor home area, i.e., $d_f \ge D_h$.
The maximum service radius, i.e., $D_{max}$,  is constrained by the physical limitation to support wireless communications. In our work, $D_{max}$ is defined by the maximum distance where the average outage probability
of an oMS is less than or equal to $O_{\max}$ while the lowest transmission rate is used.
The service radius is chosen in the range of
$d_f  \in \left[ {D_h ,D_{\max } } \right]$ to properly balance the traffic loads in macro and femtocells.

The service coverage adaptation is implemented by using a simple  MS initiated Cell Selection (MSCS) scheme.
In MSCS, the RSS threshold for femtocell association, which is referred to as $P_{cs}$, is determined,
and an MS measures the average RSS values of transmitted signals from the neighboring mBSs and fBSs.
If the MS finds some fBSs providing the average RSS larger than $P_{cs}$, the MS associates with the best
fBS among them regardless of the RSS values from the mBSs.
Therefore, the target femtocell radius $d_{f}$ is directly related to $P_{cs}$ by
\begin{equation}
\label{eq:P_cs}
P_{cs}  = {P_{f} } \left( {Z d_f} \right)^{ - \alpha },
\end{equation}
where ${P_{f} }$ is the fixed transmission power density of the fBS.
We optimize $d_{f}$ instead of $P_{cs}$ in this paper to simplify the presentation.

Then,
we formulate our parameter optimization problem
for load balancing in hybrid access femtocell networks.
The same framework can be applied for the open access femtocell networks
by setting $\beta = 0$.
Deployment of open and hybrid access femtocells
has an advantage that it can improve the performance of the macrocell users as well as the femtocell owners by transferring the traffic load in the congested macrocells to the femtocells.
Though the open access is allowed, the fMSs expect a differentiated experience
when they use their femtocells at home.
Guaranteeing the benefits of fMSs is very important to motivate the consumers
to buy and install the femtocells.
Therefore, we aim at maximizing the average performance of mMSs
by controlling $\rho$, $d_f$, and $\beta$~(hybrid access only),
while guaranteeing the relative benefits of femtocell owners.
%
%KHI:
Specifically,
it is assumed that fMSs
expect that their average femtocell throughput should be at least $M$ times larger
than the average throughput of mMSs though they allow the open access.

%KHI:
Denoting the average throughput of an fMS, mMS, and oMS by $\overline {T}_f \left( {\rho ,d_f, \beta } \right)$, $\overline {T}_m\left( {\rho ,d_f,\beta} \right)$, and $\overline {T}_o\left( {\rho ,d_f, \beta } \right)$, respectively, our optimization problem is formulated as:
\begin{xalignat}{2}
\label{eq:original_formulation}
\mathop {\max }\limits_{\rho ,d_f, \beta} & ~ \overline {T}_m  \left( {\rho ,d_f, \beta} \right)    \\
s.t. \; \notag\\
& \overline {T}_f  \left( {\rho ,d_f, \beta} \right) \ge M\overline {T}_m \left( {\rho ,d_f, \beta } \right), \label{eq:const1} \\
& \overline {T}_o  \left( {\rho ,d_f, \beta} \right) \ge K\overline {T}_m \left( {\rho ,d_f, \beta} \right), \label{eq:const_hh} \\
& D_h  \le d_f  \le D_{\max }, \label{eq:const2} \\
& 0 \le \rho\le 1, \label{eq:const3} \\
& 0 \le \beta\le 1. \label{eq:const4}
\end{xalignat}
The macrocell users might not want to associate with fBSs if
the performance is degraded by the open access with fBSs. Therefore, the
constraint~(\ref{eq:const_hh}) is additionally introduced to guarantee the minimum performance of oMSs.
Because oMSs are not the subscribed users, $K$ is configured as a value smaller than $M$, and
we basically assume that $K=1$.

Our optimization framework is a long-term parameter optimization
based on the average system-wide performance metric.
Though a short-term local resource allocation scheme might
improve the performance of the local system efficiently,
our long-term optimization is very meaningful
in the following aspects.

\begin{figure}
\begin{center}
\includegraphics[width=0.35\textwidth]{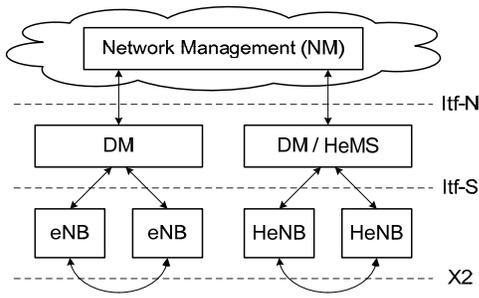}
\caption{SON architecture of 3GPP LTE system.}
\label{fig:son}
\end{center}
\end{figure}
First, the parameter optimization involving both the
macrocells and femtocells is generally performed
at a long-term interval due to the system
architecture of the two-tier cellular networks.
In 3GPP LTE system, automatic parameter configuration and optimization
are conducted based on Self Organizing Network~(SON) procedures.
Fig.~\ref{fig:son} illustrates the
SON architecture of 3GPP LTE system,
where macrocell and femtocell base stations
are referred to as Home eNodeB~(HeNB) and eNodeB~(eNB), respectively~\cite{book12Hamalainen}.
SON algorithms can be implemented either
in the end devices, i.e., HeNBs or eNBs, and/or Device Management
%PTJ: 한번 나오고 더 나오지 않는 약자
% ~(DM)
and/or Home eNodeB Management System
%PTJ: 한번 나오고 더 나오지 않는 약자
%~(HeMS)
, and/or Network Management~(NM). However, it is natural that the joint optimization of
macrocell and femtocell parameters is performed by NM in a centralized manner,
because no direct interface between eNB and HeNB exists.
Short-term parameter optimization in the centralized entity, e.g., NM,
is almost impossible due to the limited processing power and
the limited Operations, Administration and Maintenance
%PTJ: 한번 나오고 더 나오지 않는 약자
%~(OAM)
bandwidth on the interfaces among the entities.
Consequently, the joint parameter optimization involving both
macrocells and femtocells need to be performed
in the centralized entity at a long-term interval.

Second, short-term resource allocation schemes cannot generally
consider the network-wide performance
due to the excessive overhead and lack of information.
On the other hand, long-term optimization can consider
the network-wide performance thanks to the relatively small overhead
per unit time and relaxed time constraint.
Therefore, our long-term optimization is not
incompatible with but complementary to the short-term
resource allocation schemes
in the sense that the long-term optimization framework can provide good guidelines
for the parameter configuration considering the system-wide average
performance.

Finally, our long-term parameter selection schemes have
the strength in the sense that they can be utilized
in both the self-configuration and/or self-optimization phases.
Note that SON algorithms can be categorized into self-configuration and self-optimization~\cite{3gppTS36.3.9}
according to the functionality and phases.
Self-configuration presents pre-operational procedures including
the initial parameter selection.
Though an initial parameter configuration is essential, an initial parameter
selection based on the local instantaneous optimization is not generally recommended
 because the information is very insufficient and unreliable.
 On the other hand, our algorithm which does not require the instantaneous local information can be used
 for initial parameter selection. The parameters
 determined by self-configuration  may be updated by the localized self-optimization algorithms in the operational phase.
%%% 04 System Model end %%%

%%% 05 Analysis begin %%%
\section{Numerical Analysis of Two-Tier Cellular Networks}
\label{sec:analysis}
\subsection{Average Throughput of fMS}
\label{sec:femto_model}

In this section, we analyze the average throughput performance of fMSs.
Let us consider an fMS who is located at $r_f$ away from its serving fBS.
As explained in Section~\ref{sec:system_model}, we simply assume that the distribution of fBSs
follows pure SPPP in numerical analysis.
Due to the characteristics of homogeneous Poisson point process~\cite{book93kingman},
the interference measured by a typical fMS is representative of the interference
seen by all other fMSs.
Then, similarly to the SINR models in~\cite{tit0baccelli,pimrc10cheng},
the complementary cumulative distribution function~(CCDF) of an fMS's SINR is given by
\begin{xalignat}{2}
\label{eq:sinr_femto}
F_f \left( {\Gamma |r_f} \right) \buildrel \Delta \over = &~ \Pr \left[ { SINR \ge \Gamma |r_f} \right]  \notag\\
= &~ \exp \left( { - sP_N } \right)\exp \left( { - \frac{{2\pi ^2 \lambda _f Z_5^{ - 2} \left( {sP_f } \right)^{2/\alpha _5 } }}{{\alpha _5 \sin \left( {2\pi /\alpha _5 } \right)}}} \right),
\end{xalignat}
where $s = \Gamma \left( {Z_2 r_f } \right)^{\alpha _2 } P_f^{ - 1}$.
The pathloss parameters in the above equation, i.e., $Z_2$, $\alpha_2$, $Z_5$, and $\alpha_5$, are properly chosen
from Table~\ref{table:pathloss} by considering that the fMSs are always located inside the buildings in our system model.
\begin{table}
  \centering
  \caption{Pathloss parameters }\label{table:pathloss}
  \begin{tabular}{|c||c||c|}
  \hline
  Environment & Exponent & Fixed loss (dB)  \\
  \hline\hline
  Outdoor & $\alpha_1=4$  &  $Z_1^{\alpha_1} = 30\log _{10}  f_c  - 71$\\ \hline
  Indoor &  $\alpha_2=3$ & $Z_2^{\alpha_2}= 37$\\ \hline
  Outdoor-to-indoor &  $\alpha_3=4$ &  $Z_3^{\alpha_3}= 30\log _{10}  f_c - 71 + WL$\\ \hline
  Indoor-to-outdoor &  $\alpha_4=4$ & $Z_4^{\alpha_4}= 30\log _{10}  f_c - 71 + WL$\\ \hline
  Indoor-to-indoor &  $\alpha_5=4$ & $Z_5^{\alpha_5}= 30\log _{10}  f_c - 71 + 2WL$\\ \hline
  \end{tabular}
\end{table}
The detailed derivation for (\ref{eq:sinr_femto}) is given in Appendix~\ref{appendix:fMS}.

The probability density of the distance between an fMS and its serving fBS is $r_f$ is given by
$\frac{{2r_f}}{{D_h^2 }}$.
%KHI:
Hence, the average spectral efficiency of an fMS, denoted by $\overline{B}_f$, is calculated as
\begin{equation}
\label{eq:B_f_2}
\overline {B}_f   =\sum\limits_{l = 1}^L {\int_0^{D_h } {b_l \left[ {F_{f } \left( {\Gamma _l |r_f} \right) - F_{f } \left( {\Gamma _{l + 1} |r_f} \right)} \right]\frac{{2r_f}}{{D_h^2 }}dr_f} },
\end{equation}
where $b_l$, $\Gamma _{l}$, and L are the spectral efficiency of rate index $l$, the SINR threshold
to utilize the rate index, and the number of available rate sets, respectively.

At a given average spectral efficiency value,
the average throughput of an fMS is degraded when the femtocell resources
are shared with other macrocell users
i.e., oMSs,
 in the hybrid (and~open) access mode.
Let us denote the random variable for the number of oMSs who are associated
with an fBS as $N_{o}$.
Then, the average throughput $\overline {T}_f $ is given by

\begin{xalignat}{2}
\label{eq:T_f_1}
\overline {T}_f \left( {\rho  ,d_f, \beta } \right)
=& E\left[ {\beta \rho W  B_f} \right]
+ E\left[ {\frac{{(1-\beta) \rho W  B_f}}{{N_{o} + 1}}} \right] \notag\\
=& \beta\rho W \overline B _f + \left( {1 - \beta } \right)\rho W \overline B_f
 E\left[ \frac{1}{N_o + 1} \right],
\end{xalignat}
\noindent where $W$ is the system bandwidth and $B_f$ is the spectral efficiency of an fMS.
In Eq.~(\ref{eq:T_f_1}), the dedicated resources to fMS contribute to the first term while the second term is due to the shared resources between fMS and oMS.
In addition, the equality holds because the spectral efficiency of an fMS, i.e., $\overline B_f$,  is independent of the
number of $N_{o}$.
If we refer to the service area of a given femtocell as $y$,
the number of oMSs in the femtocell's service area
follows Poisson random variable with the mean $\lambda _u y$, and
the probability mass function~(pmf) of $N_{o}\left(y\right)$ is given by
\begin{equation}
\label{eq:macro_dist}
f_{N_{o}\left(y\right)}\left[ {k} \right] \sim \frac{{\left( { \lambda _uy } \right)^k {\mathop{\rm e}\nolimits} ^{ - \lambda _uy } }}{{k!}},
\end{equation}
and we obtain the expectation value as
\begin{equation}
\label{eq:N_derive}
E\left[ {  \left. {\frac{1}{{N_{o}\left(y\right)  + 1}}}  \right| y   } \right] = \sum\limits_{k = 0}^\infty  {  \frac{{\left( {\lambda _u y} \right)^k e^{ - \lambda _u y} }}{{\left(k+1\right)k!}} = \frac{{1 - e^{ - \lambda _u y} }}{{\lambda _u y}}}.
\end{equation}
Let us denote the average service area of a femtocell as $x = E\left[ y \right]$.
From~(\ref{eq:T_f_1}) and~(\ref{eq:N_derive}),
we obtain the approximated value for the average throughput of an fMS as follows:
\begin{xalignat}{2}
\label{eq:T_f_2}
\overline T _f \left( {\rho ,d_f, \beta } \right)
= &~ \beta\rho W \overline B _f + \left( {1 - \beta } \right)\rho W \overline B _f E_y \left[ {\frac{{1 - e^{ - \lambda _u y} }}{{\lambda _u y}}} \right]  \notag\\
\cong &~ \beta \rho W \overline B _f  + \frac{{\left( {1 - \beta } \right)\rho W \overline B _f \left( {1 - e^{ - \lambda _u x} } \right)}}{{\lambda _u x}}.
\end{xalignat}
The average throughput of an fMS in open access femtocell
is obtained by setting $\beta=0$.

Here, we obtain the average service area, i.e., $x$, which is a function of the target
service radius $d_f$.
A macro user associates with an mBS only when
no fBS exists within $d_f$ meters from the user.
According to the property of SPPP, the probability that
the distance $R$ between a specific mMS and its nearest fBS is larger than $r_1$ is given by
\begin{equation}
\label{eq:nearest_cdf}
\Pr \left( {R > r_1 } \right) = \Pr \left( {{\rm{No~fBS~closer~than~}}r_1 } \right) = e^{ - \pi \lambda_f r_1^2 }.
\end{equation}
Therefore, the probability that a macrocell user is covered by an fBS is given by
${1 - e^{ - \pi d_f^2 \lambda _f } }$, and
it is equivalent to the fact that the fBSs cover the area of $A_m\left( {1 - e^{ - \pi d_f^2 \lambda _f } } \right)$ on average.
Because the average number of fBSs in a macrocell area is $A_m \lambda _f$,
the average area of a femtocell region can be approximated by
\begin{equation}
\label{eq:x}
x\left( d_f \right)   = \frac{{1 - e^{ - \pi d_f^2 \lambda _f } }}{{\lambda _f }}.
\end{equation}
Because $d_f$ and $x$ have an one-to-one relationship, we use $x$
as the control parameter instead of $d_f$ in the rest of the paper for the simplicity
of presentation.

\subsection{Average Throughput of mMS}
\label{sec:macro_model}

In this section, we model the average throughput performance of an mMS,
where mMS is defined by the macrocell user who is currently associated
with the mBS.
Let us refer to the distance between an mMS and its serving mBS
as $r_m$.
From the assumption that each fBS fully covers its indoor home area,
SINR CCDF  of an
mMS for a given $r_m$ is obtained by
\begin{equation}
\label{eq:F_m}
F_{m } \left( {\Gamma |r_m} \right) = \exp \left( { - \frac{{\Gamma \left( {P_N   } \right)}}{{P_m \left( {Z_1 r_m} \right)^{ - \alpha_1 } }}} \right).
\end{equation}
As shown in Table~\ref{table:pathloss}, $Z_1$ and $\alpha_1$ are the fixed pathloss value and the pathloss exponent in outdoor environments,
respectively. The detailed derivation for the above equation is shown in Appendix~\ref{appendix:mMS}.

The  probability that a macrocell user becomes an mMS, which
is referred to as $p_{mMS}$, is given by
\begin{equation}
\label{eq:p_mms}
p_{mMS}  = e^{ - \pi \lambda _f d_f^2 }  = 1 - \lambda _f x,
\end{equation}
where $x$ is the average service area of a femtocell derived in (\ref{eq:x}).
As shown in (\ref{eq:p_mms}) that $p_{mMS}$ is  independent of the location of
the macrocell user, i.e., the distance between the user and the mBS,
due to the random distribution of fBSs.
Therefore, if we refer to the distance between an mMS and its serving mBS as $r_m$,
the probability density function~(PDF) of $r_m$ is also given by
\begin{equation}
\label{eq:f_R_m}
f_{R_m } \left( {r_m } \right) = \frac{{2r_m }}{{D_m^2 }}.
\end{equation}

From (\ref{eq:F_m}) and (\ref{eq:f_R_m}), the average spectral efficiency of an mMS is given by
\begin{equation}
\label{eq:B_m}
\overline {B}_m   =\sum\limits_{l = 1}^L {\int_0^{D_m } {b_l \left[ {F_{m } \left( {\Gamma _l |r_m} \right) - F_{m } \left( {\Gamma _{l + 1} |r_m} \right)} \right]\frac{{2r_m}}{{D_m^2 }}dr_m} },
\end{equation}
where $\Gamma _{L + 1}=\infty$.
In order to calculate the above equation, we calculate
\begin{xalignat}{2}
\label{eq:integration}
 \int_0^{D_m } {F_{m } \left( {\Gamma_l |r_m} \right)r_m} dr_m
 & = \int_0^{D_m } {r_m e^{ - \beta_{l} r_m^\alpha}} dr_m,
\end{xalignat}
where  $\beta_l$ is defined by $P_N\Gamma_l Z_{1}^{\alpha_1}  P_m^{ - 1}$.
By substituting $- \beta_{l} r^{\alpha_1}$ with $y$, the above equation is obtained by
\begin{xalignat}{2}
\label{eq:integration2}
\int_0^{\beta _l D_m^{\alpha_1}  } {\frac{1}{{\alpha_1 \beta _l }}\left( {\frac{y}{{\beta _l }}} \right)^{\frac{{2 - \alpha_1 }}{\alpha_1 }} e^{-y}} dy &= \frac{ \beta_{l} ^{ - \frac{2}{\alpha_1 }} }{\alpha_1} G \left( {\frac{2}{\alpha_1 },\beta_l D_m^{\alpha_1}  } \right),
\end{xalignat}
where $G \left( {a, b} \right) \buildrel \Delta \over = \int_0^b {t^{a - 1} e^{-t}dt}$  is the incomplete gamma function.

Although femtocell deployment does not change the average spectral efficiency of an mMS,
the average throughput of an mMS can be improved by deploying the femtocells, because
the number of mMSs sharing the macrocell resource is reduced by relocating some macrocell users, i.e., oMSs,
to the femtocells.
In our system model, the average number of macrocell users is given by
$\overline N_u = A_m \lambda _u$.
Let us refer to
the number of mMSs in a macrocell area for a given $x$ as
$N_m$.
From  (\ref{eq:p_mms}),
the expectation of $N_m$ is given by $\overline N_m \left(x\right)  = A_m \lambda _u \left( {1 - \lambda _f x} \right)$.
We approximate $N_m$ as a Poisson random variable with the mean of $\overline N_m$, and
hence, the average throughput of an mMS is given by
\begin{xalignat}{2}
\label{eq:T_m}
\overline T _m \left( {\rho ,x} \right)
& = \sum\limits_{n = 1}^\infty  {\frac{{\left( {1 - \rho } \right) W \overline B _m }}{n}} \frac{{nf_{N_m } \left[ n \right]}}{{\sum\limits_{k = 1}^\infty  {kf_{N_m } \left[ k \right]} }} \notag\\
&  \cong \frac{{\left( {1 - \rho } \right) W \overline B _m \left( {1 - e^{ - A_m \lambda _u \left( {1 - \lambda _f x} \right)} } \right)}}{{A_m \lambda _u \left( {1 - \lambda _f x} \right)}}.
\end{xalignat}

\subsection{Average Throughput of oMS}
\label{sec:oa_model}

In order to complete the load balancing problem formulated in~(\ref{eq:original_formulation}),
we analyze the average throughput of an oMSs and the maximum service radius of femtocells, i.e.,
$\overline T_{o}$ and $D_{\max}$, respectively.
When the target femtocell service radius is configured as $d_f$,
a macrocell user associates with its nearest fBS  if the
distance between the user and fBS is equal to or smaller than $d_f$.
From (\ref{eq:p_mms}), the probability that a
macrocell user becomes an oMS is given by $\lambda_fx$.
Let us refer to the distance between an oMS and its serving fBS
as $r_o$.	
Then, the conditional probability density function of $r_{o}$
is given by
\begin{equation}
\label{eq:f_R_oa}
f_{R_{o} } \left( {r_{o} |x} \right) = \left\{ {\begin{array}{*{20}c}
   {\frac{{2\pi \lambda _f r_{o} e^{ - \pi \lambda _f r_{o}^2 } }}{{\lambda _f x}},} & {0 \le r_{o}  \le  \sqrt { \frac{{\ln \left( {1 - \lambda _f x} \right)}}{{- \pi \lambda _f }}}},  \\
   {0,} & \mathrm{otherwise}.  \\
\end{array}} \right.
\end{equation}
When $r_{o}$ is given,
the SINR CCDF of an oMS is given by:
\begin{equation}
\label{eq:sinr_femto_OA_1}
\begin{array}{l}
 F_o \left( {\Gamma \left| {r_o } \right.} \right)
  \\ = \exp \left( { - \frac{{\pi \lambda _f \sqrt {sP_f } }}{{Z_I^2 }}\left( {\frac{\pi }{2} - \tan ^{ - 1} \left( {\frac{{Z_I^2 r_o^2 }}{{\sqrt {sP_f } }}} \right)} \right) - sP_N } \right), \\
 \end{array}
 \end{equation}
where
\begin{equation}
\left( {Z_I ,s} \right) = \left\{ {\begin{array}{*{20}c}
   {\left( {Z_4 ,P_f^{ - 1} \Gamma \left( {Z_4 r_o } \right)^{\alpha _4 } } \right),} & {r_o  \ge D_h, }  \\
   {\left( {Z_5 ,P_f^{ - 1} \Gamma \left( {Z_2 r_o } \right)^{\alpha _2 } } \right),} & \mathrm{otherwise.}  \\
\end{array}} \right.
\end{equation}
The detailed derivation for the above equation is found in Appendix~\ref{appendix:oaMS}.

Then, we obtain the average spectral efficiency of an oMS.
If we refer to the probability that an oMS chooses the  rate index $l$
as $f_L \left[ {l|r_{o} } \right] = F_{o} \left( {\Gamma_l |r_{o} } \right) - F_{o} \left( {\Gamma_{l + 1} |r_{o} } \right)$,
the average spectral efficiency of an oMS  is given by
\begin{equation}
\label{eq:B_o_done}
\overline B _{o} \left( x \right) = \sum\limits_{l = 1}^L {\int\limits_0^{d_f\left( x \right) } {b_l f_L \left[ {l|r_{o} } \right]f_{R_{o} } \left( {r_{o} | x} \right)dr_{o} } },
\end{equation}
where $f_{R_{o} } \left( {r_{o} | x} \right)$ is shown in (\ref{eq:f_R_oa}).

We refer to the random variable representing the number of oMSs in a
femtocell area as $N_{o}$.
As done for the average throughput analysis of an fMS
in (\ref{eq:T_f_2}), we approximately assume that $N_{o}$ is
a Poisson random variable with the mean of $\lambda_ux$.
Because $N_{o} + 1$ users including an fMS equally share
the femtocell resources in the open access mode
and only $1-\beta$ fraction of intra-femtocell resources are allowed
to the oMSs,
the average throughput of oMSs is expressed by
\begin{equation}
\label{eq:th_oa_0}
\overline T _{o} \left( {\rho ,x,\beta } \right) = \sum\limits_{n_{o} = 1}^\infty  {\frac{{\rho \left( {1 - \beta } \right) W \overline B _{o} \left( x \right)}}{{n_{o} + 1}}} \frac{{n_{o}f_{N_{o} } \left[ n_{o} \right]}}{{\sum\limits_{k = 1}^\infty  {kf_{N_{o} } \left[ k \right]} }},
\end{equation}
where
{\small
\begin{equation}
\label{eq:th_oa_1}
\sum\limits_{n = 1}^\infty  {\frac{{nf_{N_o } \left[ n \right]}}{{n + 1}}}  = \frac{1}{{\overline N _o }}\sum\limits_{n = 2}^\infty  {\left( {n - 1} \right)} \frac{{\overline N _o ^n e^{ - \overline N _o } }}{{n!}} = \frac{{\lambda _u x + e^{ - \lambda _u x}  - 1}}{{\lambda _u x}}.
\end{equation}}
From (\ref{eq:th_oa_0}) and (\ref{eq:th_oa_1}), $\overline T _{o}$ is obtained by
\begin{equation}
\label{eq:th_oa}
\overline T _{o} \left( {\rho ,x,\beta } \right) = \frac{{ \left( 1-\beta \right)  \rho W \overline B _{o} \left( x \right)\left( {\lambda _u x + e^{ - \lambda _u x}  - 1} \right)}}{{\left( {\lambda _u x} \right)^2 }}.
\end{equation}

Furthermore, we analyze the maximum service radius of a femtocell, i.e., $D_{max}$,
to complete the problem formulation in (\ref{eq:original_formulation}).
$D_{max}$ is an important parameter which determines the range of our design
parameter $d_f$ (or $x$).
As described in Section~\ref{sec:modeling}, we define $D_{max}$ as the maximum
target service radius where the average outage rate of an oMS is less than or equal to
$O_{\max}$.
We assume that an outage occurs when the
instantaneous SINR is less than the threshold for the lowest transmission rate,
i.e., $\Gamma_1$ in Table~\ref{table:rate_sets}.
From the definition, average outage rate of an oMS with the given target service radius $d_f$ is calculated by
\begin{equation}
\label{eq:D_max_1}
\overline O_{o} \left( d_f \right) = \int\limits_0^{d_f } {\left( {1 - F_o \left( {\Gamma _1 \left| {r_o } \right.} \right)} \right)f_{R_o } \left( {r_o \left| d_f \right.} \right)} dr_o,
\end{equation}
which looks very similar to~(\ref{eq:B_o_done}).
$D_{\max}$ is obtained from the equation $\overline O _o \left(  D_{\max}   \right) = O_{\max }$.
Though $D_{\max}$ cannot be given in a closed form, the near-optimal solution for~(\ref{eq:D_max_1}) can
easily be obtained by using the simple binary search algorithm because
$\overline O _o \left( d_f \right)$ is a monotonically increasing function of $d_f$.
%PTJ:
Because any mobile stations which are in the femtocell coverage should attached to femtocell, SINR of the oMS is likely to worse as the service radius of a femtocell, $d_f$, is larger when it is larger than threshold, i.e., the radius that a reference signal received power from the nearest fBS and mBS are about the same.
The detailed description for the binary search algorithm is omitted due to the space limitation.

%%% 05 Analysis end %%%

%%% 06 Optimization Open begin %%%
\section{Optimization in Open Access Femtocell Networks}
\label{sec:optimization_open}

\subsection{Optimization of Parameters in Open Access Femtocell Networks}
\label{subsec:optimization}

In this section, we prove that our optimization problem in open access mode becomes
a single variable convex problem in some typical environments, and the optimal parameters are obtained.
In this section, $\beta=0$ because we consider the open access femtocells.
If we define $\overline {t}_m  \left( {x } \right)$ and $\overline {t}_{fo}  \left( {x } \right)$ as
\begin{equation}
\label{eq:small_th_m}
\overline {t}_m \left( x \right) = \frac{{\overline T _m \left( {\rho ,x,\beta=0} \right)}}  {W\left({1 - \rho }\right)}
\end{equation}
and
\begin{equation}
\label{eq:small_th_fo}
\overline {t}_{fo} \left( x \right) = \min \left( {\frac{{\overline {T} _f \left( {\rho ,x,\beta=0} \right)}}{M W \rho },\frac{{\overline T _o \left( {\rho ,x,\beta=0} \right)}}{K W \rho }} \right)
\end{equation}
the optimization problem~(\ref{eq:original_formulation}) is rephrased by
\begin{xalignat}{2}
\label{eq:revised_formulation}
\mathop {\max }\limits_{\left( {\rho ,x} \right)} & ~ \left( 1-\rho \right) \overline {t}_m  \left( {x } \right)  \\
s.t. \; \notag\\
& \rho \overline {t}_{fo}  \left( {x } \right) \ge  \left( 1-\rho \right) \overline{t}_m  \left( {x } \right), \label{eq:const1_r} \\
& X_{\min} \le x \le  X_{\max}, \label{eq:const2_r} \\
& 0 \le \rho\le 1,
\label{eq:const3_r}
\end{xalignat}
where $X_{\min} = x\left( {D_h } \right)$ and $X_{\max} = x\left( {D_{\max} } \right)$ from (\ref{eq:x}).

\newtheorem{prop}{Proposition}
\begin{prop}\label{prop:optimal_bw}
The optimal $\rho^{*}$ which maximizes the objective in~(\ref{eq:revised_formulation})  is given by
%KHI:
$\rho^*  = \frac{{ \overline{t}_m \left( {x^* } \right) }}{{ \overline{t}_{fo} \left( {x^* } \right)  + \overline{t}_m \left( {x^* } \right) }}$,
where $x^{*}$ is the optimal value of $x$.
\end{prop}
\begin{IEEEproof}
See Appendix~\ref{appendix:proof1}.
\end{IEEEproof}

As inferred by (\ref{eq:small_th_fo}) and (\ref{eq:const1_r}), the performance of our load balancing
algorithm is always limited by the throughput requirement of an fMS if
$K\overline T _f \left( {\rho ,x, 1 } \right) \le M\overline T _o \left( {\rho ,x, 1} \right)$
for all $x \in \left[ {   X_{\min}, X_{\max}  } \right]$.
We define such cases by the fMS's requirement-limited environments.
Then, the following proposition holds.
\begin{prop} \label{prop:convex}
In the fMS's requirement-limited environments, the optimization
problem~(\ref{eq:revised_formulation}) is a convex optimization problem.
\end{prop}
\begin{IEEEproof} See Appendix~\ref{appendix:proof2}.
\end{IEEEproof}
From the above proposition, the optimal solution of $x$ can be efficiently obtained by using the
standard methods used for solving the convex optimization~\cite{book04boyd} if
the given environment is the fMS's requirement-limited environment.
We expect that the two-tier cellular networks mostly  operate in the fMS's requirement-limited
environments, because the oMS's throughput requirement is easily satisfied with the
reasonably large fMS's benefit requirement, i.e., $M$.
However, if the given environment is not the fMS's requirement-limited environment,
we should obtain the near-optimal solution of $x$ using the inefficient exhaustive search algorithm.
To check whether a given environment is the fMS's requirement-limited environment,
the following proposition could be useful.
\begin{prop} \label{prop:convex_condition}
If we define  $C\left( x \right) \buildrel \Delta \over = \frac{{ \left( {1 - e^{ - \lambda _u x} } \right)}}{{\lambda _u x}}$ and
$D\left( x \right) \buildrel \Delta \over = \frac{{ \left( {\lambda _u x + e^{ - \lambda _u x}  - 1} \right)}}{{\left( {\lambda _u x} \right)^2 }}$, $\frac{{\overline B _f C\left( {X_{\min}} \right)}}{{\overline B _o \left( {X_{\max}} \right)D\left( { X_{\min}} \right)}} \le \frac{M}{K}$
is the sufficient condition that a given environment is the fMS's requirement-limited environment.
\end{prop}
\begin{IEEEproof}
See Appendix~\ref{appendix:proof3}.
\end{IEEEproof}

From the (near) optimal value $x^*$, the service area control is implemented by
setting $P_{cs}$ according to~(\ref{eq:P_cs}) and~(\ref{eq:x}).
The optimal $\rho^*$ is configured according to Proposition~\ref{prop:optimal_bw}.

The optimal solution $x^*$ has the following property.
\begin{prop} \label{prop:D_max_optimal}
$\frac{{\overline{N}_{f} \overline {B}_f  }}{{M\overline {B}_m  }} > 1$, where $\overline{N}_{f}$ is the average number of fBSs in a macrocell area, is a sufficient condition that the optimal solution of the problem~(\ref{eq:revised_formulation}) is given by
$x^*  = X_{\max}$ in the typical fMS's requirement limited environments where
the average number of users in a macrocell is larger than that in a single femtocell.
\end{prop}
\begin{IEEEproof}
See Appendix~\ref{appendix:proof4}.
\end{IEEEproof}

Proposition~\ref{prop:D_max_optimal} can be interpreted as follows.
The expansion of the femtocell service area is encouraged if the average spectral efficiency of
an fMS is much larger than that of an mMS,
and the expansion of the femtocell service area is also encouraged if there are many femtocells in
the system because the sum of performance gain in the system is approximately proportional to the number of femtocells.
On the other hand, large benefit requirements from
fMSs, i.e., $M$, can limit the service area expansion to guarantee the performance of femtocell owners who are the premium users.

\subsection{Extension Considering Interference Thinning (Thin)}
\label{sec:interference_management}

As described in Proposition~\ref{prop:D_max_optimal}, the optimal performance of our load balancing
scheme is determined by the physical limitation $D_{\max}$ in many cases.
Therefore, we expect that the efficiency of our load balancing scheme can be enhanced
if  the maximum coverage of femtocell is increased by applying an interference thinning  scheme,
which reduces the interference by limiting the resource usage in fBSs while all the BSs fully utilize the given resources in our basic model.
In our interference thinning scheme,
the probability  that an fBS uses a specific resource block is limited to $\theta$, to reduce the interference among femtocells.
The channel condition is improved by using $\theta < 1$, but using small $\theta$ can reduce the aggregate throughput by decreasing chances for packet transmissions.
Our proposed scheme including the interference thinning scheme is referred to as OA-Thin.
The objective and constraints of OA-Thin is the same as those of the original problem formulation in OA scheme,
but we optimize the new control parameter, i.e., $\theta$,
as well as $\rho$ and $d_f$  to  maximize our objective while satisfying the constraints.

Some parameters in our analytic model are updated by considering $\theta$.
The SINR distribution of an fMS and an oMS, i.e., $F_f \left( {\Gamma |r_m } \right)$ in (\ref{eq:sinr_femto})
and $F_o \left( {\Gamma |r_o } \right)$ in (\ref{eq:sinr_femto_OA_1}), are updated by replacing
$\lambda_f$ in the equations with $\theta\lambda_f$.
Furthermore, $D_{\max}$ in is also updated considering $\theta$.

It is difficult to show that finding the optimal solution of OA-Thin scheme is a convex problem
in the fMS's requirement-limited environments as described in Proposition~\ref{prop:convex}.
However, if $\theta$ is given, Proposition~\ref{prop:convex} holds, and
we can find the optimal $x^{*}\left( \theta \right)$ and $\rho^{*}\left( \theta \right)$ efficiently
in the typical environments where the operation is dominated by the benefit requirement
of fMSs.
Therefore, we  repeatedly solve the convex optimization problems with various candidate $\theta$ values,
and $\theta$ which minimizes the objective function is chosen as the suboptimal $\theta^{*}$.

\subsection{Analysis of Optimal Parameters}
\label{sec:optimal}
\begin{figure*}
%CCC 121027 The units of y-axis of Fig. 5(b) & 5(c) are wrong. Since they are unit-less, (m) should be removed.
\centering \subfigure[Optimal femtocell coverage ($d_f^*$).]{
\includegraphics[width=0.3\textwidth]{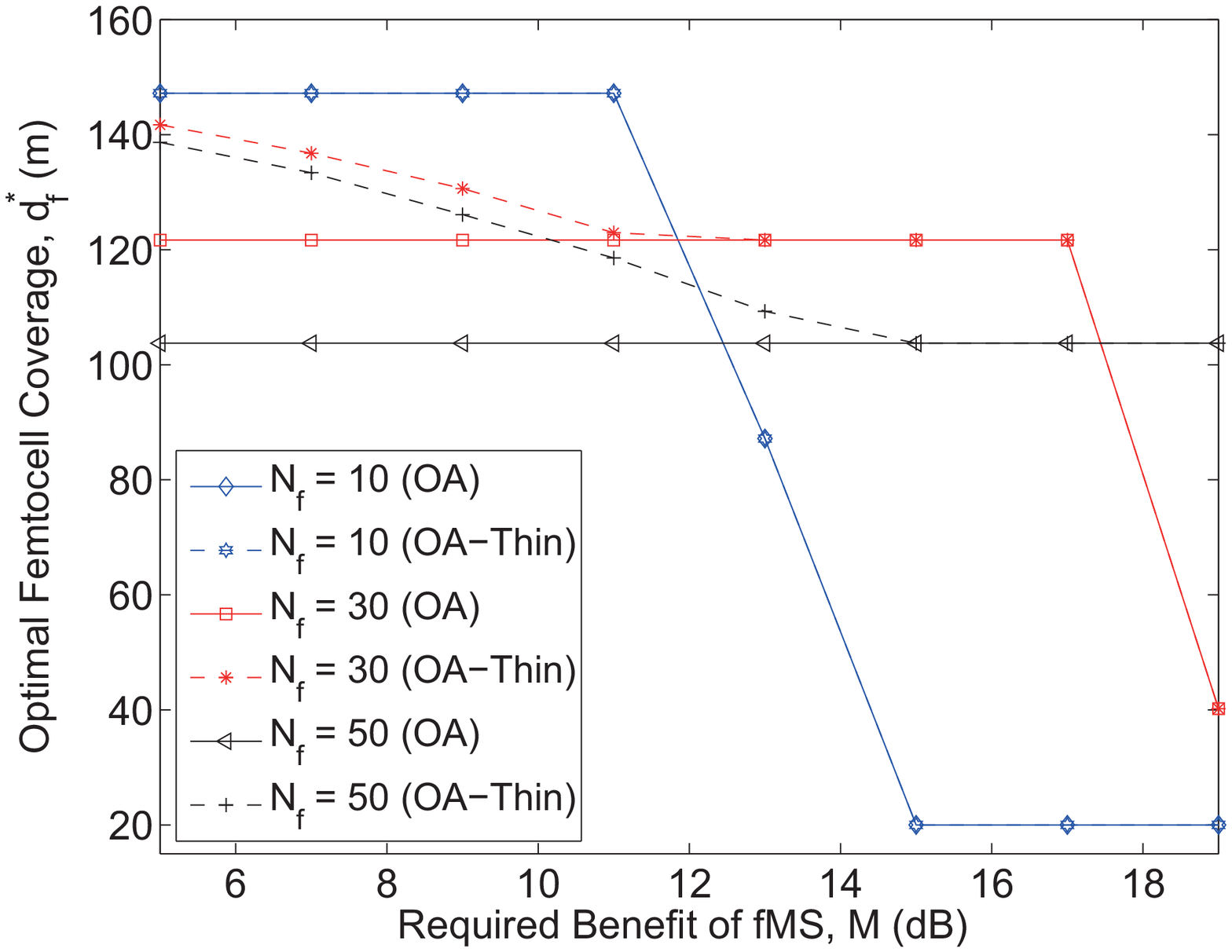}\label{fig:d_f}}\hfill
\subfigure[Optimal resource utilization ratio ($\theta^*$).]{
\includegraphics[width=0.3\textwidth]{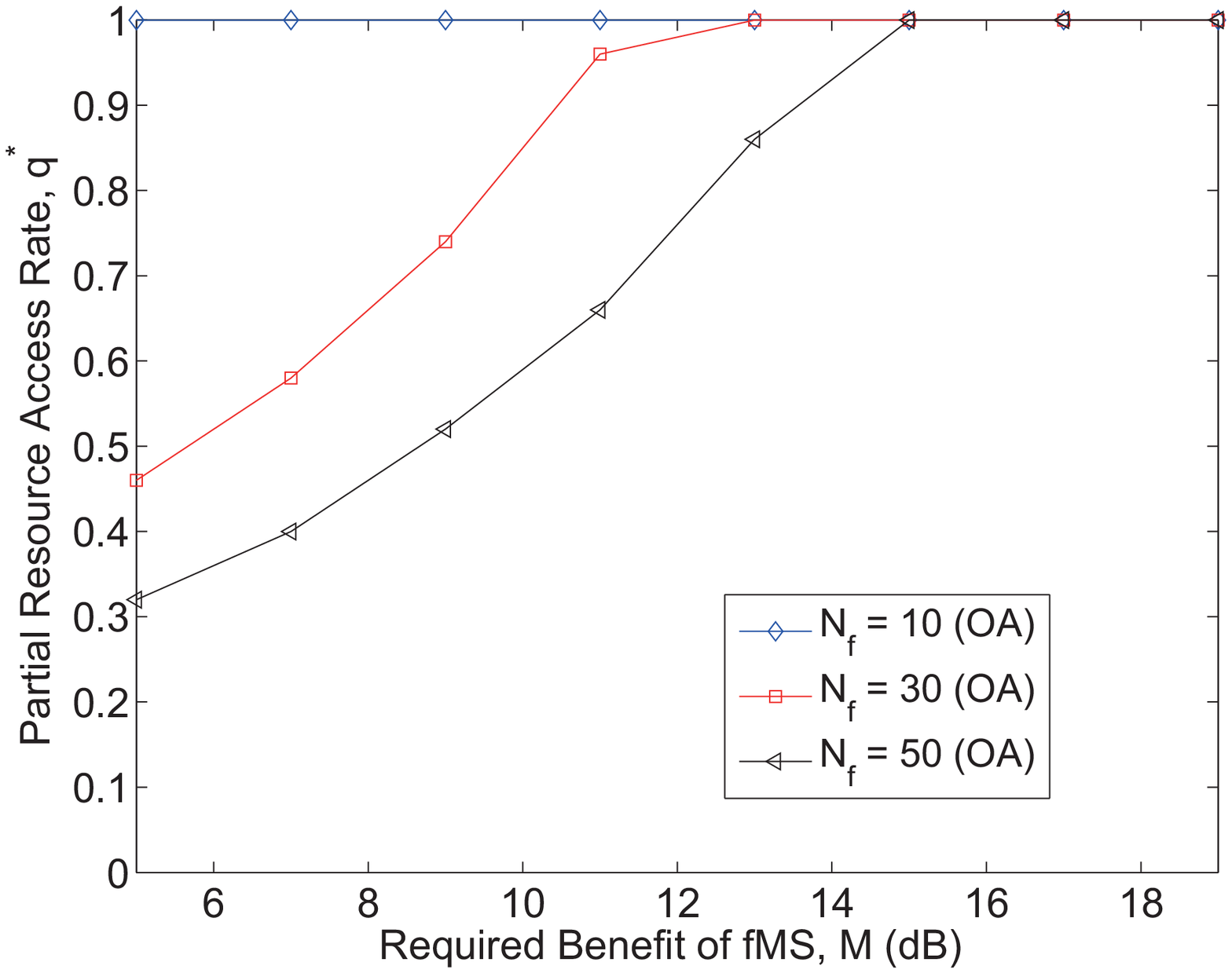}\label{fig:theta}}\hfill
\subfigure[Optimal amount of resources dedicated to femtocells ($\rho^*$).]{
\includegraphics[width=0.3\textwidth]{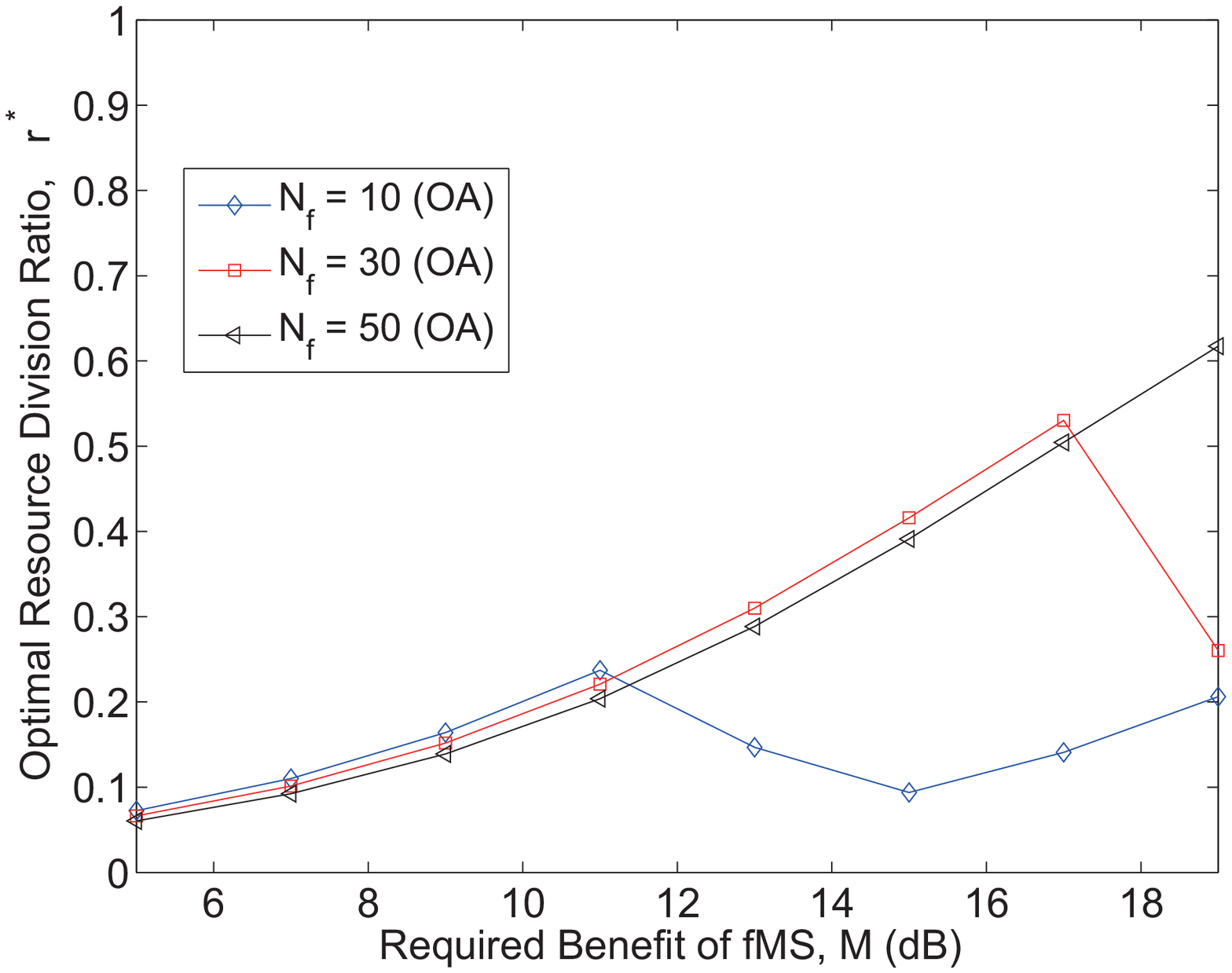}\label{fig:w_f}}
\caption{Optimal parameters.} \label{fig:optimal_parameters}
\end{figure*}

In this section, we show the optimal parameters of the proposed schemes based on our analytic model.
The basic parameters shown in Table~\ref{table:parameters} are used in the evaluations
unless mentioned otherwise, and we obtain the results with various $M$ values, where
fMSs require $M$ times higher average throughput than the average throughput of mMS.

Fig.~\ref{fig:d_f} shows the optimal service radius $d_f^*$ given by
the optimization of OA and OA-Thin schemes.
The straight lines in the figure represent $D_{\max}$ with the given average number of fBSs which is referred to as $\overline N_f$.
As $\overline N_f$ increases, $D_{\max}$ value decreases due to the increased interference.
For all $\overline N_f$ values, OA scheme determines to use $D_{\max}$ as the service radius of femtocells when $M$ is not very large.
If $M$ exceeds some threshold, the optimal femtocell radius becomes smaller than $D_{\max}$,
because sharing the femtocell resources with many oMSs
is not an efficient method to provide a large relative benefit to the fMSs.
In the sense of $\overline N_f$, $D_{\max}$ is preferred in the wider range of $M$ values when
$\overline N_f$ is large, because the offloading gain of using femtocells is more significant
with the large number of fBSs.
These results are the same results which are inferred by Proposition~\ref{prop:D_max_optimal}.

The optimal resource utilization ratio of femtocells, i.e., $\theta^*$, in OA-Thin scheme
is shown in Fig.~\ref{fig:theta}.
The needs for interference management is larger in the environments where many fBSs exist.
Therefore, Fig.~\ref{fig:d_f} shows that OA-Thin scheme chooses to expending the maximum service radius by applying $\theta < 1$ when there are $30$ or $50$ fBSs in average.
In the mean time, the interference thinning is not effective
when $M$ is very large, because it is difficult to satisfy the requirement of $M$ if fBSs use the partial resources in the femtocells.
Accordingly, Fig.~\ref{fig:theta} shows that the femtocells are required to fully utilize the dedicated femtocell resources when $M$ is large.

The optimal ratio of resources dedicated to femtocells, i.e., $\rho^*$, is shown in
Fig.~\ref{fig:w_f}. For the region where $d_f^*$ is fixed, $\rho^*$ proportionally increases
as $M$ increases. However, in the middle region where $d_f^*$
increases, $\rho^*$ decreases to properly maximize the average throughput of mMSs
while meeting the requirements for the fMS's performance.

%%% 06 Optimization Open end %%%

%%% 07 Optimization Hybrid begin %%%
%PTJ140225:
%word OA-MS changed to oMS
\section{Optimization in Hybrid Access Femtocell Networks}
\label{sec:optimization_hybrid}
In hybrid access femtocell networks,
we optimize $\beta$ as well as $x$ and $\rho$.
The analysis results for the open access femtocell networks in
the previous section are used in the optimization procedures
for the hybrid access femtocell networks.
\begin{prop}\label{prop:optimal_hybrid}
When the target femtocell service area $x$ is given, the optimal $\rho^{*}$ and $\beta^{*}$
which maximize the objective in (\ref{eq:original_formulation}) are given as follows:
\begin{equation}
\beta ^* \left( x \right) = \left\{ {\begin{array}{*{20}c}
   \frac{{ - C\left( x \right) + D\left( x \right)}}{{B\left( x \right) - C\left( x \right) + D\left( x \right)}},
 & D\left( x \right) \ge C\left( x \right),  \\
   0, & \mathrm{otherwise},  \\
\end{array}} \right.
\end{equation}
and
\begin{equation}
\rho ^* \left( x \right) = \left\{ {\begin{array}{*{20}c}
\frac{{A\left( x \right)}}{{\frac{{B\left( x \right)D\left( x \right)}}{{B\left( x \right) - C\left( x \right) + D\left( x \right)}} + A\left( x \right)}},
 & D\left( x \right) \ge C\left( x \right),  \\
   \frac{{A\left( x \right)}}{{D\left( x \right) + A\left( x \right)}}, & \mathrm{otherwise},  \\
\end{array}} \right.
\end{equation}
where
$A\left( x \right) \buildrel \Delta \over = \frac{{\overline B _m \left( {1 - e^{ - A_m \lambda _u \left( {1 - \lambda _f x} \right)} } \right) }}{{A_m \lambda _u \left( {1 - \lambda _f x} \right)}}$,
$B\left( x \right) \buildrel \Delta \over = \frac{{\overline B _f }}{M}$,
$C\left( x \right) \buildrel \Delta \over = \frac{{\overline B _f \left( {1 - e^{ - \lambda _u x} } \right)}}{{M\lambda _u x}}$, and
$D\left( x \right) \buildrel \Delta \over = \frac{{\overline B _{o} \left( x \right)\left( {\lambda _u x + e^{ - \lambda _u x}  - 1} \right)}}{{K\left( {\lambda _u x} \right)^2 }}$.
\end{prop}
\begin{IEEEproof}
See Appendix~\ref{appendix:proof5}.
\end{IEEEproof}
From Proposition~\ref{prop:optimal_hybrid} and the original problem formulation in (\ref{eq:original_formulation}),
the load balancing problem in hybrid access femtocell can be treated as the single
variable optimization with the control parameter $x$.
Unfortunately, the above optimization problem is not a convex optimization problem.
Therefore, we find a near-optimal solution by calculating the objective function over the
feasible region of $x$.
Because the original problem has been simplified to a single variable problem and
the feasible region of $x$ is bounded,
%{
we can find the near-optimal solution $x^*$
without excessively complex computations.
%}

\begin{figure}
\begin{center}
\includegraphics[width=0.3\textwidth]{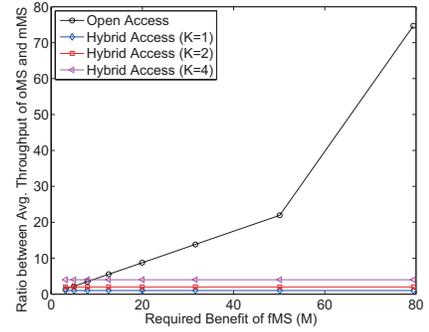}
\caption{Ratio between the average throughputs of oMS and mMS.}
\label{fig:Ratio_MK}
\end{center}
\end{figure}
The performance gain of mMS and fMS in the  hybrid access femtocell networks
is achieved at the cost of the average performance degradation of an oMS
by using $\beta>0$.
Fig.~\ref{fig:Ratio_MK} describes the ratio between the throughputs of
%KHI:
oMSs and mMSs
%OA-MSs
%
in open access femtocells and hybrid access femtocells based on analysis.
oMSs in the open access femtocells enjoy much larger throughput performance than oMSs in the hybrid access femtocells
thanks to the resource-fair intra scheduling of the fBS.
However, it is unfair that the oMSs achieve such large throughput because
the oMSs and mMSs are actually the same type of users who do not pay any cost for the femtocell deployment.
On the other hand, the hybrid access fBS distinguishes the fMS from the oMSs and the average throughput of
oMSs are properly controlled so that the similar quality of services are provided to the mMSs and oMSs.
Fig.~\ref{fig:Ratio_MK} shows that the average throughput of an oMS in  the hybrid access femtocell networks is
exactly $K$ times larger than that of an mMS, where $K$ is generally a small value.

%%% 07 Optimization Hybrid end %%%

%%% 08 Performance Evaluation begin %%%
%PTJ140225:
%word OA-MS changed to oMS
\section{Performance Evaluation}
\label{sec:evaluation}
\subsection{Evaluation Environments and Comparing Schemes}
\label{sec:eval_environments}

In this section, we evaluate our proposed schemes based on both numerical analysis and computer simulations.
As described in Section~\ref{sec:system_model}, macrocell users and fBSs are randomly deployed according
to SPPP, while one constraint that the distance between the fBSs should be less than $2D_h$
is additionally given in the simulation settings.
The channel model used for the numerical analysis and simulation is described in Section~\ref{sec:system_model}, and the basic values of the evaluation parameters are provided in Table~\ref{table:parameters}~\cite{twc10jo,itur.m.1225}.
We consider the random mobility of macrocell users in the simulations, and the performances of users located in the interested area, i.e., a single macrocell area, are considered for the performance analysis.

In our evaluations, the proposed schemes are compared with some comparing schemes.
The basic comparing scheme is CoRSSI where the mBSs and fBSs share the
same bandwidth, and a user associates with the cell which  provides the best signal strength
including both macrocells and femtocells.
CoRSSI is excellent in the aspect of sum capacity by fully reusing the
bandwidth, but the benefit achieved by
mMS can be smaller than the dedicated bandwidth based schemes
due to the limited offloading gain as will be shown in
Section~\ref{sec:simulations}.
In order to show the maximum offloading gain in a co-channel deployment, we also introduce CoLB~(Cochannel Load Balancing) scheme where
the system promotes the users to access femtocell as much as possible
while the basic requirement for service coverage
of femtocell is satisfied. The details of CoLB is described in Section~\ref{sec:simulations}.
DivRSSI is the scheme which assigns the dedicated orthogonal resources to femtocells like the proposed schemes, and each MS associates with the BS which provides the best RSSI value like CoRSSI. In DivRSSI,
the bandwidth is divided into the macro and femtocell resources with
the optimal ratio, i.e., $\rho^*$, while maintaining the constraints $\overline T_f \ge M \overline T_m$ and $\overline T_f \ge K \overline T_o$.
The optimal $\rho^*$ is chosen based on the simulation results
in DivRSSI because no numerical analysis model exists for DivRSSI.
Finally, femtocells do not allow any open or hybrid access of oMSs
to access femtocells in CoCA and DivCA, where CA stands for Closed Access.
Similarly to DivRSSI, the optimal bandwidth dedicated to femtocell is applied
for DivCA based on the simulation results, while the whole
resources are shared by macrocell and femtocell users in CoCA.

\subsection{Analysis and Simulation Results}
\label{sec:simulations}

\begin{figure*}
%CCC 121027 The units of y-axis of Fig. 5(b) & 5(c) are wrong. Since they are unit-less, (m) should be removed.
\centering \subfigure[Error between analysis and simulation results.]{
\includegraphics[width=0.3\textwidth]{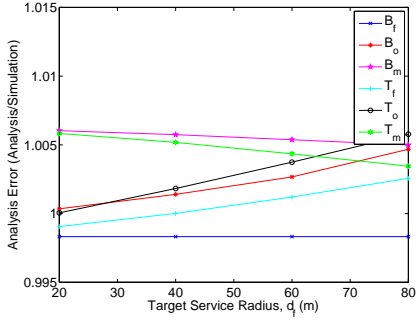}\label{fig:validation}}\hfill
\subfigure[Average throughput of mMS vs. required fMS's benefit, i.e., M.]{
\includegraphics[width=0.3\textwidth]{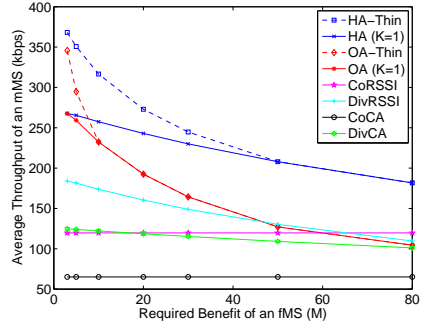}\label{fig:TH_m}}\hfill
\subfigure[Average throughput of mMS vs. number of fBSs, i.e., $N_f$.]{
\includegraphics[width=0.3\textwidth]{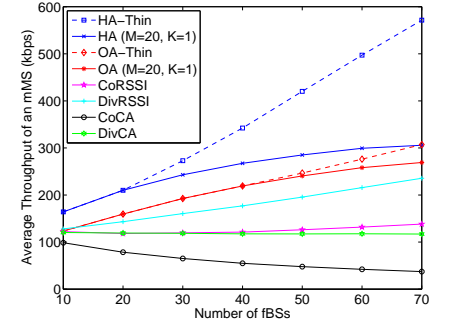}\label{fig:TH_Nf}}
\caption{Basic simulation results.} \label{fig:basic_simulations}
\end{figure*}
In this section, we evaluate the performance of proposed schemes using numerical
analysis.
The default evaluation parameters specified in Table~\ref{table:parameters}
are used unless mentioned otherwise.
First, we show that the
numerical analysis is valid in spite of the simplifying
assumptions and approximations in the numerical analysis.
We obtain the average spectral efficiency and average throughput of an fMS,
oMS, and  mMS when the service radius $d_f$ values are given.
After simulations with $10,000$ distributions, we compare the average simulation
results with the analysis result as shown in Fig.~\ref{fig:validation}.
The validation results indicate that
the errors between the results of the numerical analysis and computer simulation
are negligible, i.e., less than $1~\%$.

Fig.~\ref{fig:TH_m} shows the average throughput of an mMS.
We find that the performance of  mMSs is enhanced by utilizing the proposed
schemes in most regions. When the required benefit from fMSs is small,
the more aggressive traffic offloading from macrocell users is feasible.
Therefore, the performance gains of OA and OA-Thin are more significant when $M$ is small.
Among the comparing schemes CoRSSI provides
comparable or even better average throughput performance to mMSs than the proposed schemes when the required fMS's benefit exceeds a certain value.
This result shows that co-channel deployment could be more efficient
to guarantee very large performance benefits to fMSs.
However, this phenomenon happens when $M$ is very large, e.g., $M > 50$ in this example, and such large benefits for fMSs might not be required in reality.
HA(-Thin) schemes achieve the better average throughput of an mMS than OA(-Thin) schemes.
Especially, the gain of hybrid access femtocell is still very significant
even in the environments where $M$ is very large, while the open access femtocell's offloading gain is limited in the environments.
In HA(-Thin) schemes, the performance of mMSs and fMSs  are
improved by preventing the oMSs from achieving the performance gain
much more than necessary.
OA-Thin and HA-Thin schemes obtain the further performance gain by
improving the SINR status and extending the maximum femtocell coverage.
As discussed in Section~\ref{sec:optimal}, the partial utilization of femtocell resources, i.e., OA/HA-Thin, is preferred when $M$ is small.

The impact of the number of fBSs in a macrocell area is shown in Fig.~\ref{fig:TH_Nf}.
In the open access femtocell networks, the performance of
mMS is enhanced as the number of fBSs increases because the chances for
the offloading of macrocell's traffic increases.
As the previous propositions
implicate, the interference thinning  in OA/HA-Thin becomes useful
when the number of fBSs is
large where the load balancing efficiency is maximized.
Obviously, the performance of CoCA scheme, i.e., co-channel deployment based on the
closed access femtocells, is rapidly degraded as $N_f$ increases due to the excessive
co-channel interference.

Load balancing gain can also be enhanced
in co-channel deployment scenario by expanding
the service coverage of femtocells.
CoLB scheme expands the service coverage of femtocells
as much as possible while the basic requirement for
signal quality is satisfied.
A weight factor for femtocell access, i.e., $\delta_f$,
is notified by the system under CoLB scheme, and a macrocell user associates
with a femtocell if $\delta_f RSSI_{f,\max} > RSSI_{m}$, where
$RSSI_{f,\max}$ is the maximum received signal strength from
the nearby fBSs and $RSSI_{m}$ is the received signal strength from the mBS, respectively.
CoLB is identical to CoRSSI when $\delta_f = 1$.
Though the offloading gain obtained
in the macrocell tier increases as $\delta_f$ increases, the maximum $\delta _f$ is limited by the service quality constraint, e.g., average outage rate, of oMSs.
For a fair comparison with the proposed schemes, we use
the maximum $\delta _f$ which satisfies $\overline O _o \left( {\delta _f } \right) \le O_{\max }$, where the outage rate threshold, i.e.,  $O_{\max }$, is the same as that used in the proposed schemes to limit the maximum femtocell service radius, i.e., $D_{\max}$.

\begin{figure}
%CCC 121027 The units of y-axis of Fig. 5(b) & 5(c) are wrong. Since they are unit-less, (m) should be removed.
\centering \subfigure[Average outage rate an oMS in CoLB scheme.]{
\includegraphics[width=0.3\textwidth]{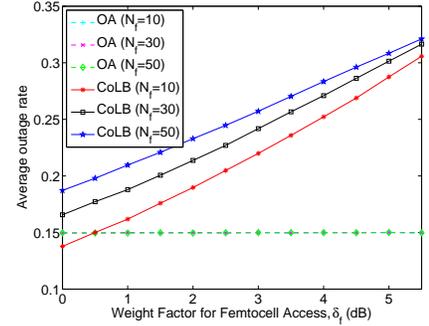}\label{fig:out_colb}}
\subfigure[Average throughput of mMS using CoLB scheme.]{
\includegraphics[width=0.3\textwidth]{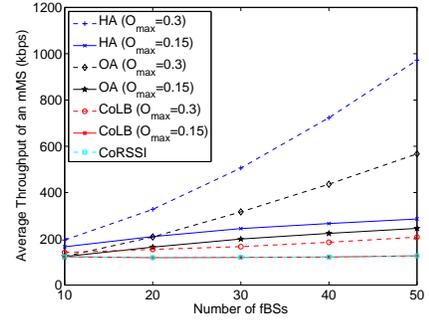}\label{fig:th_colb}}
\caption{Performance of CoLB scheme} \label{fig:colb_simulations}
\end{figure}

Fig.~\ref{fig:out_colb} shows the average outage rate of oMSs
when $\delta _f$ is given. As shown in the figure, the average outage
rate of oMSs increases as $\delta _f$ increases.
In our evaluation environments, CoLB cannot expand the service
coverage when the number of fBSs is $30$ or $50$ because the average outage
rate exceeds the configured threshold, i.e., $O_{\max} = 0.15$, even when
$\delta_f$ is $0$~dB. A small margin for coverage expansion is available
when the number of fBSs is $10$. On the other hand, the proposed scheme
adaptively manages the average outage rate according to the
number of fBSs.

Fig.~\ref{fig:th_colb} shows the average throughput
of mMS when the optimal $\delta_f$ is applied to CoLB.
To give more flexibility to coverage control, we consider
a relaxed coverage requirement where $O_{\max} = 0.3$ as well as the basic
requirement, i.e., $O_{\max} = 0.15$.
With $O_{\max} = 0.15$, the performance of CoLB is very close to that of
CoRSSI due to the limited offloading gain,
and only a small performance gain is observed when the number of fBSs is $10$.
On the other hand, a significant performance gain is provided to mMS
with the proposed schemes.
When a relaxed average outage requirement is applied, i.e., $O_{\max} = 0.3$,
CoLB achieves a larger mMS throughput than CoRSSI for most cases
at the cost of increased outage rate. However, the performance gains of mMSs
achieved by the proposed schemes in the relaxed outage requirement
are much more significant than the one achieved by CoLB.
Note that this result does not mean that the proposed scheme
is always better than the co-channel based schemes.
The proposed schemes have strength in improving
the performance of mMSs while limiting the relative
benefits of fMSs and oMSs around the planned levels.
On the other hand, in the aspect of the total system capacity,
the co-channel based scheme is more efficient.
Therefore, the choice of the resource management should
be adaptive to  various aspects, e.g.,
the system environments, status of market, consumer's characteristics,
mobile operator's policy, and so on.

\subsection{Impact of Other Environmental Parameters}
\label{sec:other}
Our basic system model used in the previous sections
includes some simplifying assumptions, e,g, uniform user distribution,
to ensure the numerical tractability.
The numerical analysis
and optimization are important in spite of the simplification
because it gives the intuition for the system performance
and optimal parameters.
However, investigation for realistic environments would be beneficial
for further understandings.
In this section, we consider some more environmental
parameters which have not been considered in the basic
numerical model,
and the impacts of the new environmental parameters
are analyzed through the simulations.

\begin{figure*}
%CCC 121027 The units of y-axis of Fig. 5(b) & 5(c) are wrong. Since they are unit-less, (m) should be removed.
\centering \subfigure[Average throughput of mMS with heterogeneous user distribution.]{
\includegraphics[width=0.3\textwidth]{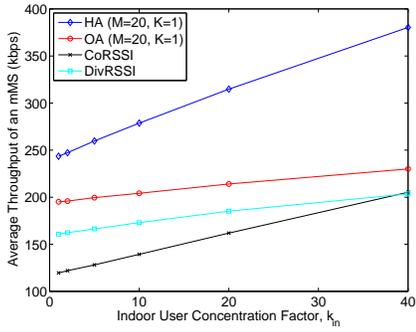}\label{fig:HETERO_TH_m_M}}\hfill
\subfigure[Limitation from the number of admissible users in an fBS.]{
\includegraphics[width=0.3\textwidth]{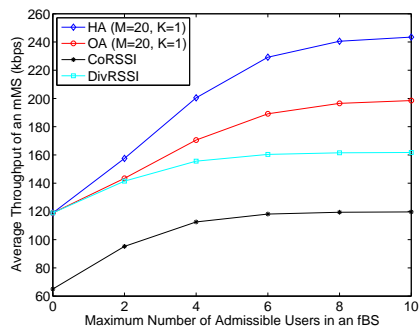}\label{fig:UE_LIMIT}}\hfill
\subfigure[Simulation results based on real deployment data.]{
\includegraphics[width=0.3\textwidth]{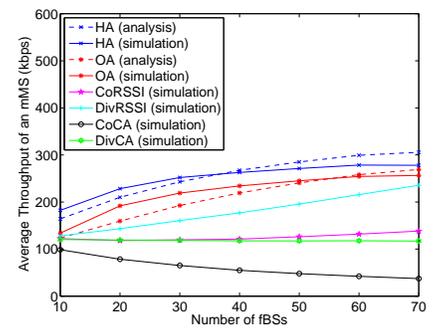}\label{fig:Seoul}}
\caption{Impact of Other Environmental Parameters.} \label{fig:other_parameters}
\end{figure*}
We assume the uniform user distribution in the basic model, but
it is generally said that indoor user density is
much larger than outdoor. Therefore, we introduce new
environmental parameter to represent the
heterogeneity for indoor and outdoor user densities, i.e.,  $k_{in}$.
We refer to the outdoor and indoor user densities as $\lambda _{u,o}$ and
$\lambda _{u,i}=k_{in}\lambda _{u,o}$, respectively.
Then, the average number of mMSs is obtained by
\begin{equation}
\label{eq:N_m_hetero}
\overline N _m  = \left( {A_m  - \lambda _f A_m x} \right)\lambda _{u,o},
\end{equation}
and the average number of oMSs in a femtocell area is given by
\begin{equation}
\label{eq:N_o_hetero}
\overline N _o  = k_{in} \lambda _{u,o} x\left( {D_h } \right) + \lambda _{u,o} \left( {x - x\left( {D_h } \right)} \right),
\end{equation}
where $x\left( {D_h } \right) = \frac{{1 - e^{ - \pi D_h^2 \lambda _f } }}{{\lambda _f }}$ from (\ref{eq:x}).
By putting (\ref{eq:N_m_hetero}) and (\ref{eq:N_o_hetero}) into
the analysis in Section~\ref{sec:analysis},
the performance of the proposed schemes
with a heterogeneous user distribution can  be numerically analyzed.

Fig.~\ref{fig:HETERO_TH_m_M} shows the average throughput of mMS
with a heterogeneous user distribution.
As shown in Fig.~\ref{fig:HETERO_TH_m_M}, the load balancing gain of all the
open access schemes increase as
the indoor user density increases because indoor femtocells
can efficiently offload the traffic of the indoor users.
The relative gain of OA scheme is reduced when the indoor user density is high,
because the comparing schemes also enjoy high offloading gain
from indoor macrocell users. On the other hand, HA scheme still
maintains the relative gain for mMS by limiting
the performance of oMSs.

In the real environments, the maximum number of users served
in a single femtocell is limited due to the capability
of the cheap femtocell device. We refer to
the maximum number of users instantaneously allowed
to access a femtocell as $N_{\max}$.
We investigate the
impact of $N_{\max}$ using  simulations.
In the simulations, each macrocell user is dropped in
a random location, and it first tries to choose the best femtocell or
macrocell based on the criterion of the employed target cell selection scheme.
If the target cell is a femtocell and the number of users in the cell
already exceeds $N_{\max}$, then the user tries to choose the next best cell
according to the cell selection criterion.
Multiple simulations are conducted for the quantized target femtocell service area candidates, i.e., $x$. Then, the optimal bandwidth division ratio and intra-femtocell
resource dedicated ratio at a given $x$, i.e., $\rho^*\left(x \right)$ and $\beta^*\left(x \right)$, are numerically obtained by using Proposition~\ref{prop:optimal_bw} and Proposition~\ref{prop:optimal_hybrid}.
Finally, the near optimal $x^*$ is chosen by comparing
the results for various $x$ values.

Fig.~\ref{fig:UE_LIMIT} shows the average throughput of an mMS
when the number of admissible users is limited by a specified value.
If only a small number of users can be served by an fBS,
the average throughput of an mMS cannot be enhanced much although
the load balancing schemes are used.
In our evaluation environments,
the offloading gain sharply decreases when the number of admissible users is less than $6$.
%CCC 121027 I added number '8' below.
On the other hand, if the number of admitted users is larger than $8$, the performance
enhancement is almost saturated.

\label{sec:building}
Our basic assumption for femtocell distribution
is that the buildings~(or houses) are
randomly distributed according to SPPP.
In order to show the performance in
the realistic environments,
we show the simulation results based on the real
deployment information for WiFi APs.
Because the femtocell service is in an early phase,
we use the location information of WiFi APs registered
in~\cite{wigle}, where we assume
that some of the WiFi APs are replaced by femtocells.

Simulation results shown in Fig.~\ref{fig:Seoul} have been obtained
based on the deployment information around the Seoul station, which
is the downtown of Seoul, Korea. In order to compare the results with
the basic results in Section~\ref{sec:simulations}, we assume that only the predefined ratio
of WiFi APs in the area are replaced by femtocells, and the density of the femtocell in
the macrocell area is known to the system. We obtain the averaged value after
$300$ simulations where different randomly chosen WiFi APs' locations
are assumed in each simulation run. The  same channel model
which is used in the numerical analysis is used for the simulations.
The meaning of this simulation result may be limited because the performance
gap between the numerical analysis and
real performance results is severely affected by the system environment.
However, the simulation results in Fig.~\ref{fig:Seoul}
shows that analysis results show similar trends to the simulation in some cases,
and the proposed schemes still have performance gain even though the
optimization based on the local information has not been performed.
Although the analysis results can be different from the performance in the real system,
the numerical analysis and optimization are still very important because it can give
the insights for the network performance in the various environments and good guidelines for the
system design.

%%% 08 Performance Evaluation end %%%

%%% 09 Conclusion begin %%%
\section{Conclusion}
\label{sec:conclusion}

In this paper, we develop the load balancing schemes which are efficient in
the environments where the open or hybrid access femtocells coexist with the macrocells.
We aim at maximizing the average throughput
performance of
mMSs while guaranteeing some amount of benefits to the
fMSs who deploy femtocells in their homes.
In order to maximize the offloading gain of the open and hybrid access femtocells, we propose
to use the separate bandwidth deployment, and we jointly optimize the
effective service areas of femtocells
and the amount of dedicated resources for femtocells.
Using the analytic model, we prove that the joint optimization problem is a convex optimization problem
in some typical environments.
We also introduce the scheme which  applies the interference thinning scheme on top of the proposed schemes
in order to further enhance  the offloading gain.
In the hybrid access femtocells, the performance of our load balancing scheme is
improved by optimally determining the amount of dedicated resources only for the fMSs.
Performance evaluation results show that the proposed schemes significantly improve the system-wide
performance while satisfying the requirements of fMSs.

%%% 09 Conclusion end %%%

\bibliographystyle{IEEEtran}
\bibliography{abrv,item}

%\newpage
%~
%\newpage

%%% 10 Appendix begin %%%

\appendices
\section{Derivation of SINR Distributions}
For the simplicity in presentation, we define a function which represents
the impact of the
aggregated interference to an MS from multiple interferers, i.e., fBSs, which are randomly located
outside the circular region with the radius of $D$.
We assume the pathloss parameters applied to the wireless links
between the MS in consideration and interfering fBSs are heterogeneous and given by
$Z$ and  $\alpha$. Then,
the aggregated interference from the multiple fBSs to the MS is expressed by
$I_a  = \sum\limits_{i \in \Lambda _{D } } {P_f \Psi _i \left( {Z r_i } \right)^{ - \alpha } }$,
where $r_i$ represents the distance between a specific interferer and the MS,
$\Lambda _{D}$ represents the set of all the interferers randomly located
outside the circular region with the radius $D$, and $\Psi$ is an independent
Rayleigh fading component of link $i$.
Let us assume that the  interferers are distributed according
to an SPPP distribution with the intensity of $\lambda_I$.
By generalizing the analysis results in~\cite{tcom11andrews}, %and Appendix A,
the Laplace transform of $I_{a}$, which is defined by
$L_{I_{a} } \left( s \right) \buildrel \Delta \over = E\left[ {e^{ - sI_{a} } } \right]$,
is calculated as follows:
\begin{equation}
\label{eq:I_out}
\begin{array}{l}
 L_{I_{a} } \left( {s|P_f ,\alpha ,Z ,\lambda _I ,D } \right)
  \\ = \exp \left( { - \pi \lambda _I Z^{ - 2} \left( {sP_f } \right)^{^{\frac{2}{{\alpha  }}} } \int\limits_{Z^2 D^2 \left( {sP_f } \right)^{ - \frac{2}{{\alpha  }}} }^\infty  {\left( {\frac{1}{{u^{\frac{{\alpha  }}{2}}  + 1}}} \right)} du} \right). \\
 \end{array}
 \end{equation}
From the integral table found in \cite{book07Gradshtein},
the Laplace transform of $I_{a}$ is given as follows in the special case that $\alpha = 4$:
\begin{equation}
\label{eq:I_out_4}
\begin{array}{l}
 L_{I_{a} } \left( {s|P_f ,4,Z ,D } \right)
  \\ = \exp \left( { - \pi \lambda _I Z^{ - 2} \sqrt {sP_f } \left( {\frac{\pi }{2} - \tan ^{ - 1} \left( {\frac{{Z^2 D^2 }}{{\sqrt {sP_f } }}} \right)} \right)} \right). \\
 \end{array}
 \end{equation}
Furthermore, in the special case that $D=0$,
\begin{equation}
\label{eq:I_out_0}
 L_{I_{a} } \left( {s|P_f ,\alpha ,Z ,0} \right)  = \exp \left( { - \frac{{2\pi ^2 \lambda _I Z^{ - 2} \left( {sP_f } \right)^{2/\alpha  } }}{{\alpha  \sin \left( {2\pi /\alpha  } \right)}}} \right).
 \end{equation}

\subsection{SINR distribution of an fMS}
\label{appendix:fMS}
Although our actual system model does not allow the deployment of an fBS
if the distance between the fBS and other fBS is less than $2D_h$,
pure SPPP distribution is assumed in the numerical analysis for
numerical tractability.
Therefore, it is assumed that the interfering fBSs are randomly
located according to SPPP with the density of $\lambda_f$ and
the indoor-to-indoor pathloss parameters are applied for the
interference links from the other fBSs.
If we refer to the interference received by a typical fMS
as $I_{f}$, the Laplace transform of $I_{f}$ is given by
\begin{equation}
\label{eq:L_fms}
L_{I_{f} }\left( s\right) = L_{I_{a} } \left( {s|P_f ,\alpha_5 ,Z_5 ,0} \right),
\end{equation}
where $\alpha_5$ and $Z_5$ are the indoor-to-indoor pathloss parameters, and
$L_{I_a } \left(  \cdot  \right)$ is given in (\ref{eq:I_out_0}).
SINR of a typical fMS is expressed by
$\gamma _f  = \frac{{\Psi \left( {Z_2 r_f} \right)^{ - \alpha_2 } P_f }}{{P_N  + I_{f} }}$,
where $r_f$ is the distance between an fMS and its serving fBS, $P_N$ is the noise power, and
$\Psi$ is an independent Rayleigh fading component of the link between an fMS and its serving fBS.
Furthermore, $Z_2$ and $\alpha_2$ are the indoor pathloss parameters as described in Table~3 of the main manuscript.
For the simplicity in presentation, we define $s = \Gamma \left( {Z_2 r_f } \right)^{\alpha _2 } P_f^{ - 1}$.
Then, the CCDF of an fMS's SINR is given by
\begin{xalignat}{2}
\label{eq:sinr_apx_fms_1}
 F_f \left( {\Gamma |r_f} \right) \buildrel \Delta \over = & ~  \Pr \left[ {\frac{{\Psi \left( {Z_2 r_f} \right)^{ - \alpha_2 } P_f }}{{P_N  + I_{f} }} \ge \Gamma } \right]  \notag\\
= & ~ \Pr \left[ {\Psi  \ge s\left( {P_N  + I_{f} } \right)} \right] = E_{I_{f} } \left[ {e^{ - s\left( {P_N  + I_{f} } \right)} } \right] \notag\\
= & ~ e^{ - sP_N } E_{I_{f} } \left[ {e^{ - sI_{f} } } \right] = e^{ - sP_N } L_{I_{f} } \left( s \right),
\end{xalignat}
where the third equality holds because $\Psi$ is an exponential random variable,
and the last equality comes from the definition of the Laplace transform.
From (\ref{eq:I_out_0}), (\ref{eq:L_fms}), and (\ref{eq:sinr_apx_fms_1}),
\begin{xalignat}{2}
\label{eq:sinr_apx_fms_2}
F_f \left( {\Gamma |r_f} \right) = \exp \left( { - sP_N } \right)\exp \left( { - \frac{{2\pi ^2 \lambda _f Z_5^{ - 2} \left( {sP_f } \right)^{2/\alpha _5 } }}{{\alpha _5 \sin \left( {2\pi /\alpha _5 } \right)}}} \right).
\end{xalignat}

\subsection{SINR distribution of an mMS}
\label{appendix:mMS}

From the constraint that each fBS fully covers its indoor home area,
i.e., $d_f \ge D_h$, all the mMS are located outside building,
and outdoor pathloss parameters, i.e., $Z_1$ and $\alpha_1$, are
applied for the wireless link between an mMS and its serving mBS.
Because no interference from the fBSs exists, SINR of
a typical mMS is expressed by
$\gamma _m  = \frac{{\Psi \left( {Z_1 r_m} \right)^{ - \alpha_1 } P_m }}{{P_N   }}$,
where $r_m$ is the distance between an mMS and its serving mBS.
If we define $s = \Gamma \left( {Z_1 r_m } \right)^{\alpha _1 } P_m^{ - 1}$,
the SINR CCDF of an mMS is obtained by
\begin{xalignat}{2}
\label{eq:sinr_apx_fms_0}
 F_m \left( {\Gamma |r_m} \right)  = & ~  \Pr \left[ {\frac{{\Psi \left( {Z_1 r_m} \right)^{ - \alpha_1 } P_m }}{{P_N   }} \ge \Gamma } \right]  \notag\\
= & ~ \Pr \left[ {\Psi  \ge sP_N} \right] = e^{ - sP_N }.
\end{xalignat}

\subsection{SINR distribution of an oMS}
\label{appendix:oaMS}
Let us consider a typical oMS who is located at $r_o$ away
from its serving fBS. If we refer to the
interference received by the oMS from the other fBSs as $I_o$,
$I_o$ is the sum of interference from the
fBSs which are randomly distributed
outside the circular region with the radius $r_o$ with the
density $\lambda_f$.
Therefore, the Laplace transform of $I_o$ is
given by
\begin{equation}
\label{eq:I_OA_1}
L_{I_o } \left( s \right) = L_{I_{a} } \left( {s|P_f ,\alpha_I ,Z_I ,\lambda _f , r_o } \right),
\end{equation}
where $Z_I$ and $\alpha_I$ are the pathloss parameters of the interfering links,
and $L_{I_a } \left(  \cdot  \right)$ is given in (\ref{eq:I_out}).
If we refer to the pathloss parameters of the desired link to the serving fBS
as $Z_d$ and $\alpha_d$,
the SINR of a typical fMS is expressed by
$\gamma _o  = \frac{{\Psi \left( {Z_d r_o} \right)^{ - \alpha_d } P_f }}{{P_N  + I_{o} }}$.
Similarly to (\ref{eq:sinr_apx_fms_1}), the CCDF of an oMS's SINR is
\begin{equation}
\label{eq:ccdf_oMS_1}
F_o \left( {\Gamma \left| {r_o } \right.} \right) = e^{ - sP_N } L_{I_{a} } \left( {s|P_f ,\alpha_I ,Z_I ,\lambda _f , r_o } \right),
\end{equation}
where $s = \Gamma \left( {Z_d r_o } \right)^{\alpha _d } P_f^{ - 1}$.
From the system model,
$\left( {Z_d ,\alpha _d ,Z_I ,\alpha _I } \right) = \left( {Z_4 ,\alpha _4 ,Z_4 ,\alpha _4 } \right)$
if the oMS is an outdoor user, i.e., $r_o \ge D_h$.
On the other hand, $\left( {Z_d ,\alpha _d ,Z_I ,\alpha _I } \right) = \left( {Z_2 ,\alpha _2 ,Z_5 ,\alpha _5 } \right)$ if
$r_o < D_h$. In our system model, $\alpha _I = 4$ because  $\alpha _4 = \alpha _5 = 4$.
Therefore, $L_{I_a } \left(  \cdot  \right)$ in (\ref{eq:I_out_4}) is utilized
to obtain the SINR CCDF of an oMS, and hence
\begin{equation}
\label{eq:sinr_femto_OA_apx}
\begin{array}{l}
 F_o \left( {\Gamma \left| {r_o } \right.} \right)
  \\ = \exp \left( { - \frac{{\pi \lambda _f \sqrt {sP_f } }}{{Z_I^2 }}\left( {\frac{\pi }{2} - \tan ^{ - 1} \left( {\frac{{Z_I^2 r_o^2 }}{{\sqrt {sP_f } }}} \right)} \right) - sP_N } \right), \\
 \end{array}
 \end{equation}
where
\begin{equation}
\left( {Z_I ,s} \right) = \left\{ {\begin{array}{*{20}c}
   {\left( {Z_4 ,P_f^{ - 1} \Gamma \left( {Z_4 r_o } \right)^{\alpha _4 } } \right),} & {r_o  \ge D_h, }  \\
   {\left( {Z_5 ,P_f^{ - 1} \Gamma \left( {Z_2 r_o } \right)^{\alpha _2 } } \right),} & {otherwise.}  \\
\end{array}} \right.
\end{equation}

\section{Proofs of Propositions}
\subsection{Proof of Proposition~1}
\label{appendix:proof1}
If $x$ is given by a fixed value and the value is feasible, the optimization problem
becomes a linear programming~(LP) problem with a single variable $\rho$. By rephrasing the constraint~(35) of the main manuscript, we obtain
that $0 \le \frac{{{t}_m \left( {x } \right) }}{{ {t}_{fo} \left( {x } \right)  + {t}_m \left( {x } \right) }} \le \rho  \le 1$. Because ${t}_m  \left( {x } \right) >0$, the objective is maximized when $\rho$ is the minimum value.
By the definition of $x^{*}$, the objective is maximized when  $\left(\rho, x\right) =  \left(  \frac{{{t}_m \left( {x^* } \right) }}{{{t}_{fo} \left( {x^* } \right)  +  {t_m} \left( {x^* } \right) }}  , x^*\right)$.

\subsection{Proof of Proposition~2}
\label{appendix:proof2}

In the fMS's benefit requirement limited environments,
$t_{fo}$ and $t_{m}$ are respectively given by
\begin{equation}
t_{fo} \left( x \right) = \frac{{ \overline B _f \left( {1 - e^{ - \lambda _u  x } } \right)}}{{M\lambda _u   x }}
\end{equation}
\begin{equation}
t_{m} \left( x \right) = \frac{{\overline B _m \left( {1 - e^{ - A_m \lambda _u \left( {1 - \lambda _f x} \right)} } \right)}}{{A_m \lambda _u \left( {1 - \lambda _f x} \right)}}.
\end{equation}
From Proposition~1, we can replace $\rho$ in the formulated problem in~(34) of the main manuscript
with $\frac{{{t}_m \left( {x } \right) }}{{{t}_{fo} \left( {x } \right)  +  {t_m} \left( {x } \right) }}$
Then, the objective function becomes a single-variable function, i.e.,
$\frac{{{t}_m \left( {x } \right) {t}_{fo} \left( {x } \right)  }}{{ {t}_{fo} \left( {x } \right)  +  {t}_m \left( {x } \right) }}$.
Accordingly, the optimization problem is equivalent to the following single-variable minimization problem:
\begin{xalignat}{2}
\label{eq:min_formulation}
\mathop {\min }\limits_{x}
& ~\frac{{M\lambda _u x}}{{\overline {B}_f  \left( {1 - e^{ - \lambda _u x} } \right)}} + \frac{{A_m \lambda _u \left( {1 - \lambda _f x} \right)}}{{\overline {B}_m \left( {1 - e^{ - A_m \lambda _u \left( {1 - \lambda _f x} \right)} } \right)  }}\\
s.t. \; \notag\\
& \frac{{1 - e^{ - \pi D_h^2 \lambda _f } }}{{\lambda _f }} \le x \le \frac{{1 - e^{ - \pi D_{\max }^2 \lambda _f } }}{{\lambda _f }}, \label{eq:const1_min}
\end{xalignat}
where the objective function is a reciprocal of the objective in the original problem.

It is well known that the minimization problem which has only the
linear constraints is a convex problem if the objective function is convex~\cite{book04boyd}.
We prove that~(\ref{eq:min_formulation}) is
a convex function by showing that both the first and second terms of~(\ref{eq:min_formulation}) are convex.

First, we show that the second derivative of the first term of~(\ref{eq:min_formulation}) is always positive when $x > 0$.
The second derivative of $\frac{{\lambda _u x}}{{\overline {B}_f  \left( {1 - e^{ - \lambda _u x} } \right)}}$ is calculated by
\begin{equation}
\label{eq:second_der}
\lambda _u^2 \overline {B}_f  ^{ - 1} e^{ - y} \left( {1 - e^{ - y} } \right)^{ - 3} \left[ {\left( {y + 2} \right)e^{ - y}  + y - 2} \right],
\end{equation}
where $y \buildrel \Delta \over =  \lambda _u x > 0$. Because $\lambda _u^2 \overline {B}_f  ^{ - 1} e^{ - y} \left( {1 - e^{ - y} } \right)^{ - 3}  \ge 0$ when $ y>0$,
the first term of~(\ref{eq:min_formulation}) is convex if $\left[ {\left( {y + 2} \right)e^{ - y}  + y - 2} \right] \ge 0$ when $ y>0$.
Clearly, the above inequality holds when $y \ge 2$, because both $\left( {y + 2} \right)e^{ - y}$ and $y - 2$ are positive
when $y \ge 2$.
When $0<y<2$, we need to prove that $e^{ - y}  \ge \frac{{2 - y}}{{2 + y}}$,
and it is satisfied if $\frac{d}{{dy}}\left[ {e^{-y} } \right]  \ge \frac{d}{{dy}}\left[ 	\frac{{2 - y}}{{2 + y}} 	 \right]$ because $ \left[ {e^{-y} } \right]_{y=0}  =  \left[ 	\frac{{2 - y}}{{2 + y}} 	\right]_{y=0}$. It is equivalent to prove that $e^y  \ge \frac{{\left( {y + 2} \right)^2 }}{4}$ in the region.
We complete the proof by showing that the following three inequalities hold:
$\left[ {e^y } \right]_{y = 0}  \ge \left[	\frac{{\left( {y + 2} \right)^2 }}{4}	\right]_{y = 0}$,
$\frac{d}{{dy}}\left[ {e^y } \right]_{y = 0}  \ge \frac{d}{{dy}}\left[ 	\frac{{\left( {y + 2} \right)^2 }}{4} 	 \right]_{y = 0}$,
and $\frac{{d^2 }}{{dy^2 }}\left[ {e^y } \right]_{y \ge 0} \ge \frac{{d^2 }}{{dy^2 }}\left[ {\frac{{\left( {y + 2} \right)^2 }}{4}} \right]_{y \ge 0}$, where the detailed calculations are omitted.
%}

Second, we prove that the second term of~(\ref{eq:min_formulation}) is also convex.
We define a new parameter $z = A_m \left( {1 - \lambda _f x} \right)$.
Then, the second term of~(\ref{eq:min_formulation}) can be expressed by $\frac{{\lambda _u z}}{{\overline B _m \left( {1 - e^{ - \lambda _u z} } \right)}}$, which
is the same form as the first term of~(\ref{eq:min_formulation}).
According to the above derivations in this appendix, $\frac{{\lambda _u z}}{{\overline B _m \left( {1 - e^{ - \lambda _u z} } \right)}}$ is a convex function of $z$.
Because an affine mapping preserves the convexity~\cite{book04boyd}, the second term of~(\ref{eq:min_formulation})
is also a convex function of $x$.

\subsection{Proof of Proposition~3}
\label{appendix:proof3}

If we define  $C\left( x \right) \buildrel \Delta \over = \frac{{ \left( {1 - e^{ - \lambda _u x} } \right)}}{{\lambda _u x}}$ and
$D\left( x \right) \buildrel \Delta \over = \frac{{\left( {\lambda _u x + e^{ - \lambda _u x}  - 1} \right)}}{{\left( {\lambda _u x} \right)^2 }}$,
the average throughputs of an fMS and an oMS are respectively given by
$\overline T _f \left( {\rho ,x} \right) = \rho W \overline B _f C\left( x \right)$ and
$\overline T _o \left( {\rho ,x} \right) = \rho W \overline B _o \left( x \right)D\left( x \right)$.
Then, the condition that a given environment is the fMS's requirement-limited environment is expressed by
$\frac{{\overline B _f C\left( x \right)}}{{\overline B _o \left( x \right)D\left( x \right)}} \le \frac{M}{K}$
for all $x \in \left[  X_{\min} ,    X_{\max}   \right]$.
If the conditions $\frac{d}{{dx}}\left[ {\overline B _o \left( x \right)} \right] \le 0$ and
$\frac{d}{{dx}}\left[ {\frac{{D\left( x \right)}}{{C\left( x \right)}}} \right] \ge 0$ are satisfied
in the all feasible $x$,
$\frac{{\overline B _f C\left( {X_{\min } } \right)}}{{\overline B _o \left( {X_{\max } } \right)D\left( {X_{\min } } \right)}} \le \frac{M}{K}$ is a sufficient condition that  a given environment is the fMS's requirement-limited environment because
\begin{equation}
\frac{{\overline B _f C\left( x \right)}}{{\overline B _o \left( x \right)D\left( x \right)}} \le \frac{{\overline B _f }}{{\overline B _o \left( {X_{\max } } \right)C^{ - 1} \left( {X_{\min } } \right)D\left( {X_{\min } } \right)}} \le \frac{M}{K}.
%\frac{{\overline B _f C\left( x \right)}}{{\overline B _o \left( x \right)D\left( x \right)}} \le \frac{{\overline B _f C\left( {X_{\min } } \right)}}{{\overline B _o \left( {X_{\max } } \right)D\left( {X_{\min } } \right)}} \le \frac{M}{K}
\end{equation}

Therefore, we complete the proof by showing that $\frac{d}{{dx}}\left[ {\overline B _o \left( x \right)} \right] \le 0$
and $\frac{d}{{dx}}\left[ {\frac{{D\left( x \right)}}{{C\left( x \right)}}} \right] \ge 0$.
Firstly, it is trivial that $\frac{d}{{dx}}\left[ {\overline B _o \left( x \right)} \right] \le 0$  because the average transmission
rate of oMSs should be decreased by expanding the service radius of the femtocells.
Secondly, we show that $\frac{d}{{dy}}\left[ {\frac{{D\left( {y/\lambda _u} \right)}}{{C\left( {y/\lambda _u} \right)}}} \right] \ge 0$ which is equivalent to
$\frac{d}{{dx}}\left[ {\frac{{D\left( x \right)}}{{C\left( x \right)}}} \right] \ge 0$, where $y = \lambda_u x$.
We need to show that $\frac{d}{{dy}}\left[ {\frac{{e^{ - y}  + y - 1}}{{y\left( {1 - e^{ - y} } \right)}}} \right] \ge 0$
in the whole feasible region of $y$, and the condition is equivalent to
\begin{equation}
\label{eq:54}
\frac{5}{4} \ge \left( {e^{ - y}  - \frac{1}{2}} \right)^2  + e^{ - y} \left( {y - 1} \right)^2,
\end{equation}
and the above inequality holds if both $\frac{1}{4} \ge \left( {e^{ - y}  - \frac{1}{2}} \right)^2$
and $e^{ y} \ge  \left( {y - 1} \right)^2$ are satisfied in $y \ge 0$.
It is trivial that $\frac{1}{4} \ge \left( {e^{ - t}  - \frac{1}{2}} \right)^2$ because
$0 \le e^{ - y}  \le 1$ when  $y \ge 0$.
Furthermore, $ e^{ y} \ge \left( {y - 1} \right)^2$ when
%CCC Is my change correct?
$0 \le y \le 1$ because $e^y  \geq 1$ and $\frac{d}{{dy}}\left( {y - 1} \right)^2  < 0$ in the region.
When $y \ge 1$, it can be shown that $ e^{ y} \ge \left( {y - 1} \right)^2$
because the following three inequalities hold:
$\left[ {e^y } \right]_{y = 1}  \ge \left[ {\left( {y - 1} \right)^2 } \right]_{y = 1}$,
$\frac{d}{{dy}}\left[ {e^y } \right]_{y = 1}  \ge \frac{d}{{dy}}\left[ {\left( {y - 1} \right)^2 } \right]_{y = 1}$, and
$\frac{{d^2 }}{{dy^2 }}\left[ {e^y } \right]_{y = 1}  \ge \frac{{d^2 }}{{dy^2 }}\left[ {\left( {y - 1} \right)^2 } \right]_{y = 1}$.
%}

\subsection{Proof of Proposition~4}
\label{appendix:proof4}
We define $f\left( t \right) \buildrel \Delta \over = \frac{t}{{1 - e^{ - t} }}$ and
its first derivative as $f^{'} \left( t \right)$.
Then, the first derivatives of the first and second terms
of (\ref{eq:min_formulation}) are given by
$\frac{{M\lambda _u }}{{\overline B _f }}f^{'} \left( t \right)\left| {_{t = \lambda _u x} } \right.$
and
$ - \frac{{\lambda _u \lambda _f A_m }}{{\overline B _m }}f^{'} \left( t \right)\left| {_{t = A_m \lambda _u \left( {1 - \lambda _f x} \right)} } \right.$, respectively.
Hence, if
$\frac{{\lambda _f A_m }}{{\overline B _m }} = \frac{{\overline{N}_{f} }}{{\overline B _m }} > \frac{M}{{\overline B _f }}$
and
$f^{'} \left( {A_m \lambda _u \left( {1 - \lambda _f x} \right)} \right) \ge f^{'} \left( {\lambda _u x} \right)$
always hold in the feasible region of $x$, the maximum cell coverage is optimal.
According to the derivation in Appendix A,
$f^{'} \left( t \right)$ is an increasing function of $t$.
Accordingly, $f^{'} \left( {A_m \lambda _u \left( {1 - \lambda _f x} \right)} \right) \ge f^{'} \left( {\lambda _u x} \right)$
always holds if $A_m \lambda _u \left( {1 - \lambda _f x\left( {D_{\max } } \right)} \right) \ge \lambda _u x\left( {D_{\max } } \right)$.
The last condition represents that the average number of users in a macrocell is larger than
the average number of users in a single femtocell with the maximum cell coverage, and the condition is satisfied
in the general environments. Therefore, $\frac{{\overline{N}_{f} \overline {B}_f  }}{{M\overline {B}_m  }} > 1$ is a sufficient condition that the optimal femtocell coverage is the maximum value in the general environments.

\subsection{Proof of Proposition~5}
\label{appendix:proof5}
The objective function of our load balancing scheme is not a function of $\beta$.
Instead, $\beta$ only influences the feasible region of the other control parameters
$\left( {\rho ,x} \right)$ determined by the constraints.
From~(4) and~(5) of the main manuscript, the constraint related to $\beta$ is given by
\begin{equation}
\label{ineq:min}
\min \left( {\frac{{\overline T _f \left( {\rho ,x,\beta } \right)}}{M},\frac{{\overline T _{o} \left( {\rho ,x,\beta } \right)}}{K}} \right) \ge \overline T _m \left( {\rho ,x} \right).
\end{equation}
Because the right side of the above inequality is the objective function of our optimization problem,
$\beta$ which maximizes the left side of the above inequality
is an (or one of) optimal parameter(s) which maximizes the objective function of
the original problem, i.e.,
\begin{equation}
\label{eq:beta_opt_0}
\begin{array}{l}
 \beta ^* \left( {\rho ,x} \right) = \mathop {\arg \max }\limits_\beta  \,\,\,\,\min \left( {\frac{{\overline T _f \left( {\rho ,x,\beta } \right)}}{M},\frac{{\overline T _{o} \left( {\rho ,x,\beta } \right)}}{K}} \right), \\
 \,\,\,\,\,\,\,\,\,\,\,\,\,\,\,\,\,\,\,\,\,\,\,\,\,\,\,\,s.t.\,\,\,0 \le \beta  \le 1. \\
 \end{array}
 \end{equation}
At a given $\left( {\rho ,x} \right)$, we refer to the solution of $\beta$ which solves the equation
$\frac{{\overline T _f \left( {\rho ,x,\beta } \right)}}{M} = \frac{{\overline T _{o} \left( {\rho ,x,\beta } \right)}}{K}$ as $Q$.
It is easy to show that ${\overline T _f \left( {\rho ,x,\beta } \right)}$ and ${\overline T _{o} \left( {\rho ,x,\beta } \right)}$
are monotonically increasing and decreasing functions of $\beta$, respectively.
Due to the monotonicity of the two functions, the solution real value $Q$ always exists, and
$\min \left( {\frac{{\overline T _f \left( {\rho ,x,\beta } \right)}}{M},\frac{{\overline T _{o} \left( {\rho ,x,\beta } \right)}}{K}} \right)$
 is an increasing function when
$\beta  < Q$ and a decreasing function when $\beta \ge  Q$.
Therefore, the optimal parameter $\beta^*$ maximizing (\ref{eq:beta_opt_0})
in the range of $\beta  \in \left[ {0,1} \right]$ is given by
\begin{equation}
\label{eq:beta_opt_1}
\beta ^*  = \left\{ {\begin{array}{*{20}c}
   {Q,} & {0 \le Q \le 1},  \\
   {0,} & {Q < 0},  \\
   {1,} & {Q > 1}.  \\
\end{array}} \right.
\end{equation}
From the definition of $A\left( x \right)$, $B\left( x \right)$
$C\left( x \right)$, and $D\left( x \right)$ in Proposition~5,
$\frac{{\overline T _f \left( {\rho ,x,\beta } \right)}}{M} = W \rho \beta \left( {B\left( x \right) - C\left( x \right)} \right) + W \rho C\left( x \right)$ and $\frac{{\overline T _{o} \left( {\rho ,x,\beta } \right)}}{K} = W \rho \left( {1 - \beta } \right)D\left( x \right)$.
Accordingly, $Q$ is calculated by
\begin{equation}
\label{eq:Q}
Q = \frac{{D\left( x \right) - C\left( x \right)}}{{B\left( x \right) + D\left( x \right) - C\left( x \right)}}.\\\
\end{equation}
Because  ${B\left( x \right)}$, ${C\left( x \right)}$, and ${D\left( x \right)}$ are positive in
the region of interest, $Q \le 1$.
From (\ref{eq:beta_opt_1}) and (\ref{eq:Q}), we obtain the optimal $\beta^*$ as a function of $x$ as follows:
\begin{equation}
\beta ^* \left( x \right) = \left\{ {\begin{array}{*{20}c}
   \frac{{ - C\left( x \right) + D\left( x \right)}}{{B\left( x \right) - C\left( x \right) + D\left( x \right)}}
 & D\left( x \right) \ge C\left( x \right)  \\
   0 & otherwise,  \\
\end{array}} \right.
\end{equation}

%{
By using $\beta ^*$, the load balancing problem in hybrid access femtocell networks
can be rephrased by inserting
\begin{equation}
t _{fo }\left( x \right)  = \left\{ {\begin{array}{*{20}c}
   {\frac{{B\left( x \right)D\left( x \right)}}{{B\left( x \right) + D\left( x \right) - C\left( x \right)}},} & {D\left( x \right) \ge C\left( x \right),}  \\
   {D\left( x \right),} & {otherwise,}  \\
\end{array}} \right.
\end{equation}
into (35)~of the main manuscript.
Because Proposition~1 still holds,
\begin{equation}
\rho ^* \left( x \right) = \left\{ {\begin{array}{*{20}c}
\frac{{A\left( x \right)}}{{\frac{{B\left( x \right)D\left( x \right)}}{{B\left( x \right) - C\left( x \right) + D\left( x \right)}} + A\left( x \right)}},
 & D\left( x \right) \ge C\left( x \right),  \\
   \frac{{A\left( x \right)}}{{D\left( x \right) + A\left( x \right)}}, & otherwise.  \\
\end{array}} \right.
\end{equation}
%}

%ddd reduce - MOVE TO APPENDIX

\section{Discussion for Orthogonal Deployment and Its Applications}
\label{sec:orthogonal}

Generally, co-channel deployment of femtocells might be preferred because the
spectral efficiency can be maximized by the co-channel deployment.
However, we in this paper propose a load balancing scheme in
two-tier cellular networks based on the orthogonal channel deployment,
and the evaluation results have shown that
the orthogonal channel deployment can be beneficial in some aspects.
The advantages of the orthogonal
channel deployment inferred by this paper are summarized as follows:
\begin{enumerate}
\item Orthogonal deployment can enhance the maximum service coverage of femtocells by removing the cross-tier interference from macrocells, and it is beneficial to increase the amount of macrocell load transferred %macrocell load
    to femtocells. Many macrocell users can be served by open and hybrid access femtocells in the orthogonal deployment.
\item The service quality provided to each type of users can be flexibly controlled. Therefore, the mobile operators can adaptively select the system parameters to provide the `adequate' service quality for each type of users based on its own policy. The `adequate' service quality can be very different according to the deployment environment, market status, and characteristics of end consumers.
%CCC 121027 I simply removed the following incomplete sentence since we don't have enough space anyway.
%For example, the end consumers ????
\end{enumerate}
Considering orthogonal deployment's capability for no cross-tier interference and
the large coverage provisioning, orthogonal deployment can be beneficial in the following conditions:	
\begin{enumerate}
\item The traffic load in the macrocell is so high that large traffic offloading from macrocell area is essential, and/or
\item Service area is not perfectly covered by macrocell area so that the femtocells are expected to help the service coverage extension, and/or
\item The fMSs do not require the large performance benefit from using femtocells so that it is not necessary to allocate the whole resources to fMSs and oMSs as co-channel deployment based schemes do, and/or
\item The end consumers are not willing to spend much money to use the femtocell service so that the financial benefits obtained by solely femtocell selling is not expected to be very large, and/or
\item The cost reduction due to the reduced macrocell deployment or increased benefit obtained from the enhanced macrocell user performance is expected to be significant.
\end{enumerate}

\ifCLASSOPTIONcaptionsoff
  \newpage
\fi

\end{document}